\documentclass[aps,prd,twocolumn,nofootinbib,superscriptaddress]{revtex4-2}
\usepackage{graphicx}  
\usepackage{dcolumn}   
\usepackage{bm}        
\usepackage{amsmath,amssymb,amsxtra,bm,epsfig}   
\usepackage{float}
\usepackage[dvipsnames]{xcolor}
\usepackage{xspace}
\usepackage{cancel,soul,ulem}

\usepackage{tikz-feynman}
\usepackage{stmaryrd}
\tikzfeynmanset{compat=1.1.0}
\usetikzlibrary{arrows}

\tikzset{	
	vertex/.style={circle,draw, minimum size=1.5em},	
	edge/.style={->,> = latex'}	
}
\usepackage[bookmarks, breaklinks, colorlinks,urlcolor=blue, citecolor=red, 
linkcolor=blue]{hyperref}
\hypersetup{
	colorlinks=true,
	linkcolor=red,
	filecolor=magenta,
	urlcolor=ForestGreen,
	citecolor=RoyalBlue}
	
\usepackage[]{caption}

\newcommand{\be}{\begin{eqnarray*}}
	\newcommand{\ee}{\end{eqnarray*}}

\newcommand{\bee}{\begin{eqnarray}}
	\newcommand{\eee}{\end{eqnarray}}
\newcommand{\beeq}{\begin{equation}}
	\newcommand{\eeq}{\end{equation}}

\usepackage{xspace}

\newcommand{\ba}{\begin{array}}
	\newcommand{\ea}{\end{array}}
\newcommand{\bd}{\begin{displaymath}}
	\newcommand{\ed}{\end{displaymath}}
\newcommand{\besub}{\begin{subequations}}
	\newcommand{\eesub}{\end{subequations}}
\newcommand{\bea}{\begin{eqnarray}}
	\newcommand{\eea}{\end{eqnarray}}

\def\q2 {q^2}

\usepackage{tikz}
\usetikzlibrary{arrows,shapes}
\usetikzlibrary{trees}
\usetikzlibrary{matrix,arrows} 				
\usetikzlibrary{positioning}				
\usetikzlibrary{calc,through}				
\usetikzlibrary{decorations.pathreplacing}  
\usepackage{pgffor}							

\usetikzlibrary{decorations.pathmorphing}	
\usetikzlibrary{decorations.markings}
\tikzset{
	vector/.style={decorate, decoration={snake}, draw},
	provector/.style={decorate, decoration={snake,amplitude=2.5pt}, draw},
	antivector/.style={decorate, decoration={snake,amplitude=-2.5pt}, draw},
	fermion/.style={draw=black, postaction={decorate},
		decoration={markings,mark=at position .55 with {\arrow[draw=black]{>}}}},
	fermionbar/.style={draw=black, postaction={decorate},
		decoration={markings,mark=at position .55 with {\arrow[draw=black]{<}}}},
	fermionnoarrow/.style={draw=black},
	gluon/.style={decorate, draw=black,
		decoration={coil,amplitude=4pt, segment length=5pt}},
	scalar/.style={dashed,draw=black, postaction={decorate},
		decoration={markings,mark=at position .55 with {\arrow[draw=black]{>}}}},
	scalarbar/.style={dashed,draw=black, postaction={decorate},
		decoration={markings,mark=at position .55 with {\arrow[draw=black]{<}}}},
	scalarnoarrow/.style={dashed,draw=black},
	electron/.style={draw=black, postaction={decorate},
		decoration={markings,mark=at position .55 with {\arrow[draw=black]{>}}}},
	bigvector/.style={decorate, decoration={snake,amplitude=4pt}, draw},
}

\tikzstyle{block} = [draw, rectangle, 
minimum height=3em, minimum width=6em]

\begin{document}

\title{Exploring Leptogenesis in the Era of First Order Electroweak Phase Transition}

\author{Dipendu Bhandari}
\email{dbhandari@iitg.ac.in}
\affiliation{Department of Physics, Indian Institute of Technology Guwahati, Assam-781039, India}

\author{Arunansu Sil}
\email{asil@iitg.ac.in}
\affiliation{Department of Physics, Indian Institute of Technology Guwahati, Assam-781039, India}

\begin{abstract} We present a novel approach for implementing baryogenesis via leptogenesis at low scale within neutrino seesaw framework where a sufficient lepton asymmetry can be generated via out-of-equilibrium CP-violating decays of right handed neutrinos (RHNs) even when their mass falls below the Standard Model (SM) Higgs mass. It becomes possible by keeping the sphaleron in equilibrium below its conventional decoupling temperature $T_{\rm sp}^{\rm SM} \sim131.7$ GeV in SM so as to facilitate the conversion of lepton asymmetry to baryon asymmetry at such a low scale, thanks to the flexibility of the bubble nucleation temperature in case the electroweak phase transition (EWPT) is of first order. The scenario emerges as an exciting (and perhaps unique) possibility for low scale leptogenesis, particularly if the Universe attains a reheating temperature lower than 131.7 GeV. We show that a stochastic gravitational wave, characteristic of such first order EWPT, may be detected in near future detectors, while the presence of RHNs of mass as low as 35 GeV opens up an intriguing detection possibility at current and future accelerator experiments. 

\end{abstract}
\maketitle
Understanding the baryon asymmetry of the Universe (BAU)~\cite{Planck:2018vyg} remains a core problem in particle physics and cosmology. Among various proposals, generating a lepton asymmetry from the out-of-equilibrium decay of heavy Standard Model (SM) singlet right-handed neutrinos (RHNs) $N_i$ into SM lepton ($l_L$) and Higgs ($H$) doublets in the early Universe, called leptogenesis~\cite{Luty:1992un,Fukugita:1986hr,Plumacher:1996kc,Covi:1996wh}, and transferring it to the baryon sector via sphaleron process ~\cite{Kuzmin:1985mm,Arnold:1987mh,Arnold:1987zg,Bochkarev:1987wf,Khlebnikov:1988sr} manifest itself as the most natural explanation for BAU. This is particularly because of its close connection to the neutrino mass generation mechanism via Type-I seesaw~\cite{Minkowski:1977sc,Gell-Mann:1979vob,Mohapatra:1979ia,Yanagida:1979as,Schechter:1980gr,Schechter:1981cv}. This minimal SM extension not only takes care of expounding the tiny neutrino mass and BAU but also provides candidate for another unsolved conundrum~\cite{Datta:2021elq}, the dark matter \cite{1967ApJ...148..175J,SDSS:2003eyi}.

In thermal leptogenesis, the mass of the decaying RHN is typically constrained by the Davidson-Ibarra bound~\cite{Davidson:2002qv} $M_N \gtrsim 10^9$ GeV for hierarchical RHNs. Alongside, with quasi-degenerate RHNs, resonant enhancement~\cite{Flanz:1994yx,Flanz:1996fb,Pilaftsis:1997jf,Pilaftsis:2003gt} of lepton asymmetry production helps to achieve leptogenesis with lighter RHN masses around electroweak (EW) scale $\mathcal{O}(160 ~{\rm{GeV}})$~\cite{DOnofrio:2014rug,DOnofrio:2015gop} and above, making such scenarios testable at current and future colliders. The apparent possibility to enhance such detection prospect to next level by lowering the RHN mass scale further down 
is limited by the fact that the sphalerons decouple below $T_{\rm sp}^{\rm SM} = 131.7$ GeV in SM~\cite{DOnofrio:2014rug}, preventing lepton-to-baryon asymmetry conversion.

Leptogenesis can also proceed via Higgs decay~\cite{Hambye:2016sby} (considering thermal effects) with (quasi-degenerate) RHNs mass below the EW scale, in a typical temperature window between $T_{\rm sp}^{\rm SM}$ to EW symmetry breaking temperature $T_{\rm EW}\sim \mathcal{O}(160 ~{\rm{GeV}})$. Additionally, leptogenesis via oscillations \cite{Akhmedov:1998qx} remains viable with RHNs having mass as low as in the GeV regime, occurring at temperatures well above $T_{\rm sp}^{\rm SM}$. Recently, we also showed~\cite{Bhandari:2023wit} that temperature-dependent RHN masses can yield sufficient lepton asymmetry at high temperatures, even if their zero-temperature masses lie in the GeV regime.

In all scenarios involving sub-EW mass RHNs, leptogenesis requires the temperature of the Universe to exceed $T_{\rm sp}^{\rm SM}$. Again, any asymmetry (lepton or baryon) must arise in a post-inflationary epoch (otherwise, a complete erasure is inevitable), the onset of which is usually (with instantaneous reheating) marked by the reheating temperature $T_{\rm RH}$. The lower bound on $T_{\rm RH}$ being few MeV only (from BBN), an interesting and relevant question emerges: is it feasible to realize leptogenesis when the reheating temperature lies below the $T_{\rm sp}^{\rm SM}, ~i.e.$ 131.7 GeV? In such scenarios, with the Universe never reaching $T_{\rm sp}^{\rm SM}$, lepton-to-baryon asymmetry conversion is inhibited. Hence, this intriguing possibility remains unexplored in the literature to the best of our knowledge.

In this work, we demonstrate that leptogenesis indeed remains a possibility at such a low scale ($i.e.$ even below  $T_{\rm sp}^{\rm SM}$) provided the EW phase transition (EWPT) is of strongly first order\footnote{Note that for the above mentioned low scale leptogenesis scenarios, the EWPT is considered to be a smooth cross-over.} having a characteristic bubble nucleation temperature $T_n$, at which one bubble on an average is nucleated per horizon. Although bubbles of broken phase (inside which Higgs $vev$, $\langle H \rangle = v(T) \neq 0$) begin to form at a certain critical temperature $T_c$ above $T_n$, they can't grow beyond a critical size and initiate the conversion of the Universe from the symmetric phase ($\langle H \rangle =0$) to the broken one until the temperature drops to $T_n$. Hence, the symmetric phase is prevalent in the entire Universe till $T_n$. Note that $T_n$ can be kept below $T_{\rm sp}^{\rm SM}$ in scenarios with the first order EWPT (FOEWPT) in general. This may have interesting consequence in terms of leptogenesis as in such a circumstance, a RHN having mass $M_N$ within a range $T_n<M_N<T_{\rm sp}^{\rm SM}$ can decay out of equilibrium ($N \rightarrow l_L + H$) and produces a lepton asymmetry in the Universe (in symmetric phase) that can still be converted into BAU via sphalerons. This is due to the fact that sphalerons (above $T_n$ and below 131.7 GeV) remaining in the symmetric phase are in equilibrium as the expansion of the broken phase bubbles (beyond a critical size) pervading the Universe are yet to be started. Immediately below $T_n$, tunneling to the true vacuum ($v (T) \neq 0$) from the false one starts to proceed efficiently, leading to the nucleation of bubbles. During the nucleation, the baryon asymmetry (produced outside the bubble) is engulfed inside with the gradual expansion of the bubble. As the sphaleron rate is exponentially suppressed: $\Gamma_{\rm sph}^{\rm (in)} \sim T^4 {\rm exp}\left[-8\pi v(T)/g_{\rm w} T \right]$~\cite{Cline:2006ts} within the true vacuum bubbles, sphalerons decouple immediately and hence, the enclosed baryon asymmetry remains preserved.

Note that the pivotal role in materialising the above idea is played by the nucleation temperature $T_n$, associated to FOEWPT, being smaller than the sphaleron decoupling temperature of the SM. This can in general be possible within any framework of FOEWPT, making the proposal model independent. Hence, before going to evaluate the parameter space specific to such a possibility, we specify here the background of estimating the $T_n$ first. The FOEWPT occurs through the nucleation of true-vacuum ($\langle H \rangle = v(T) \neq 0$) bubbles. Although broken-phase bubbles can begin to form at the critical temperature $T_c$, at which false ($\langle H \rangle =0$) and true minima become degenerate, they can't grow beyond a critical size and initiate the conversion of the Universe from the symmetric phase ($\langle H \rangle =0$) to the broken one as their nucleation remain suppressed due to the small false-vacuum decay rate. 
Hence, the symmetric phase is prevalent in the entire Universe till the temperature at which the probability of forming at least one bubble per horizon volume reaches order one, and the transition from the false to the true vacuum effectively proceeds, initiating bubble nucleation. The characteristic temperature at which this occurs is known as the nucleation temperature ($T_n$), which can be determined using the relation~\cite{Guth:1981uk,Guth:1979bh}
\begin{equation}
N_b(T_n)= \int_{T_n}^{T_c} \frac{dT}{T}\frac{\Gamma(T)}{{\mathcal{H}}(T)^4}=1, \label{bubble-number}
\end{equation}
where $N_b$ is the number of bubbles per horizon, and $\Gamma(T) \simeq T^4 \left( \frac{S_3}{2\pi T} \right)^{3/2} {\rm exp}(-S_3/T)$ is the false vacuum decay rate~\cite{Coleman:1977py,Linde:1980tt,Linde:1981zj} with $S_3$ being the 3-dimensional Euclidean action for $O(3)$-symmetric bounce solution, computed numerically using the package {\bf FindBounce}~\cite{Guada:2020xnz} (see appendix~\ref{ap:2}) and ${\mathcal{H}}(T) = 0.33 \sqrt{g_*}\frac{T^2}{M_p}$ ($M_p = 2.4 \times 10^{18}$ GeV is the reduced Planck mass) is the Hubble parameter. Though the exact nucleation temperature depends on the specific model parameters of the extensions of the SM (responsible for FOEWPT) such as scalar~\cite{Espinosa:2011ax,Profumo:2007wc} or fermionic extensions~\cite{Davoudiasl:2012tu}, or via the inclusion of a non-renormalizable dimension-6 operator to the SM Higgs potential~\cite{Chala:2018ari,Bodeker:2004ws,Delaunay:2007wb,Grojean:2004xa,Huang:2015izx,Huang:2016odd}, it is in general plausible for $T_n$ to remain significantly lower, even below $T_{\rm sp}^{\rm SM}$. In the following, as a case study, we adopt a minimal setup by incorporating a dimension-6 operator to the SM Higgs potential to analyze the FOEWPT quantitatively and show to what extent $T_n$ can be lowered in such a picture. Based on such finding, we plan to enter estimating low scale leptogenesis connected to a $T_n$ below the $T_{\rm sp}^{\rm SM}$ to set the lower bound of RHN mass for successful leptogenesis.

To accommodate the proposal ascribed above, where EWPT needs to be strongly first order, we must go beyond the SM since within the SM, the EWPT remains a smooth crossover with the Higgs mass 125 GeV. From the minimality point of view, we therefore extend the SM Higgs doublet potential by introducing a non-renormalizable dimension-6 operator ${(H^{\dagger}H)^3}/{\Lambda^2}$ as proposed in various studies~\cite{Chala:2018ari,Bodeker:2004ws,Delaunay:2007wb,Grojean:2004xa,Huang:2015izx,Huang:2016odd}, 
where $\Lambda$ is the cut-off scale and $H^T = \left[ G_1 + i G_2, \phi + h + i G_3\right]/{\sqrt{2}}$ with $G_i$ as the Goldstone bosons and $h$ the physical Higgs field. Here, $\phi$ corresponds to the background (classical) field, the tree-level potential of which is given by
\begin{equation}
V_{(\phi)}^{\rm tree}= -\frac{\mu_h^2}{2}\phi^2 + \frac{\lambda_h}{4}\phi^4 + \frac{1}{8}\frac{\phi^6}{\Lambda^2}. \label{tree-potential}
\end{equation}
In order to analyze the phase transition precisely, it is essential to incorporate the one-loop finite temperature correction $V_T$ into it, which results in (detailed in appendix~\ref{ap:1}) an effective potential $V_{\rm eff} =  V_{(\phi)}^{\rm tree} + V_{T}(\phi,T)$ for $\phi$ as 
\begin{equation}
V_{\rm eff} =  \left( -\frac{\mu_h^2}{2}+\frac{1}{2}c_h T^2 \right)\phi^2 + \left( \frac{\lambda_h}{4} + \frac{\lambda_1}{4}T^2 \right) \phi^4 + \frac{1}{8}\frac{\phi^6}{\Lambda^2}, \label{high-T-potential}
\end{equation}
where
\begin{equation*}
c_h=\frac{1}{16}\left( \frac{4 m_{h}^2}{v_0^2} + 3 g_{\rm w}^2 + g_Y^2 +4 y_t^2 -12 \frac{v_0^2}{\Lambda^2} \right), \hspace{0.28 cm} \lambda_1=\frac{1}{\Lambda^2}, \label{ch-lamda1-value}
\end{equation*}
with $m_h = 125$ GeV and $v_0$= 246 GeV as the Higgs mass and Higgs $vev$ at zero-temperature respectively. At this stage, we restrict ourselves with terms upto order $T^2$ and hence, daisy diagrams~\cite{Espinosa:1992gq} are not incorporated. The one-loop Coleman-Weinberg correction being negligible compared to the thermal correction, is not included also. Note that we adopt such approximated form of the effective Higgs potential, Eq.~\ref{high-T-potential}, to provide a simplified illustration of how the upper and lower bound of $\Lambda$ can be obtained and how the lowest $T_n$ for successful first-order electroweak phase transition can be determined. However, in the subsequent part of analysis to demonstrate the phenomena of bubble nucleation and the transition of space volume from the false vacuum to the true vacuum phase, we employ the full effective thermal potential, including daisy resummation (as described in appendix~\ref{ap:1}) to accurately estimate the final parameter space of our study.


As seen from the effective potential in Eq.~ \ref{high-T-potential}, two degenerate minima form at the critical temperature $T_c$ (corresponding to a particular $\Lambda$), separated by a potential barrier in between. The parameter $\Lambda$ controls the barrier height between the local and global minima, as detailed in appendix~\ref{ap:1}. Increasing $\Lambda$ diminishes the barrier height, weakening the first-order nature of the transition. Employing the full effective thermal potential in the {\bf CosmoTransition} package~\cite{Wainwright:2011kj}, we find that for $\Lambda > 810$ GeV, the transition is no longer first-order. Thereby, $\Lambda = 810$ GeV serves as the upper limit. Conversely, a lower bound on $\Lambda$ can be exercised by analyzing the dynamics of bubble nucleation. 

\begin{figure}[!htb]
	\includegraphics[width=1\linewidth]{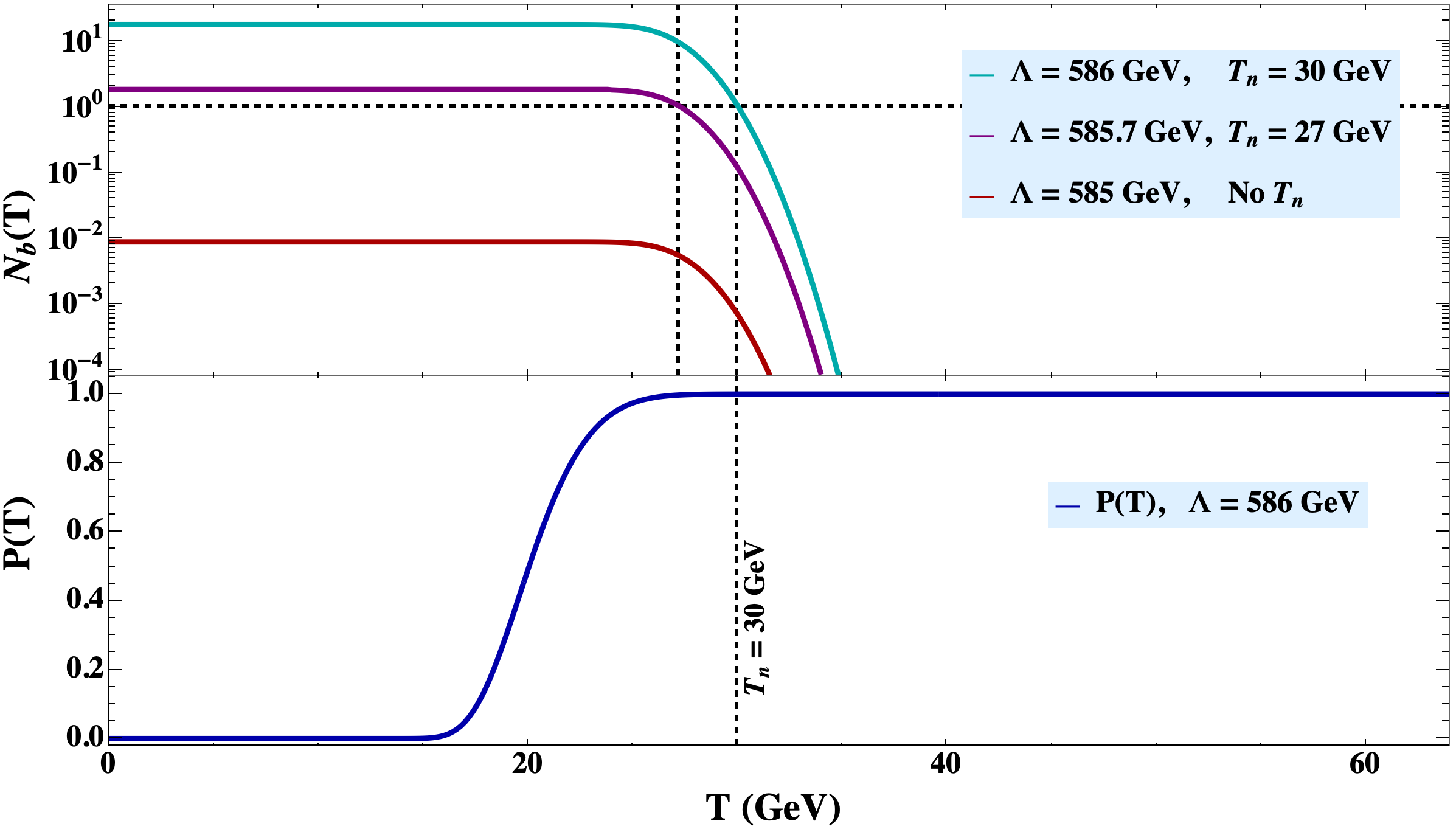}
	\caption{Number of bubbles per horizon as function of $T$.}
	\label{fig:bubble-number-plot}
\end{figure}
In Fig.~\ref{fig:bubble-number-plot} upper panel, we illustrate the $N_b(T)$ dependence on temperature, for three closely spaced values of $\Lambda$ where the corresponding $T_n$ values (as obtained from Eq. \ref{bubble-number} using the high-temperature approximated potential in Eq.~\ref{high-T-potential}) are indicated by the vertical black dashed lines. It turns out that for $\Lambda=585$ GeV, $N_b$ never reaches 
	$\mathcal{O}(1)$, indicating no bubble nucleation and hence an incomplete phase transition. In contrast, for $\Lambda \gtrsim 585.7$ GeV, the condition $N_b \geq 1$ is satisfied, revealing that the false-vacuum decay rate surpasses the expansion rate of the universe, $\it{i.e.}$, $\Gamma(T) \gtrsim {\mathcal{H}}(T)^4$ at a specific $T_n$ depending on $\Lambda$. 
	
Upon incorporating the full effective thermal potential, including both the Coleman–Weinberg and daisy corrections, the above analysis establishes the lower limit $\Lambda = 571.6$ GeV. Combining this with the earlier upper bound from barrier height perspective, the viable range of $\Lambda$ and correspondingly $T_n (\Lambda)$ (and $T_c$ too) for a strongly FOEWPT is
\begin{equation} 
 571.6 ~{\rm{GeV}} \lesssim \Lambda \lesssim 810 ~{\rm{GeV}}; ~~33.8 ~{\rm{GeV}} \lesssim T_n \lesssim 119 ~{\rm{GeV}}, \label{eq:Lambda-limit}
 \end{equation}
where the criteria for successful bubble nucleation and the completion of the phase transition are satisfied.


Observing that $T_n$ ($T_c$) can be as low as 34 (79) GeV for a strongly FOEWPT, we now explore the implications of such low $T_n$ on the sphaleron decoupling. Since the sphaleron rate remains unsuppressed in an environment of vanishing the SM Higgs $vev$ ($\it{i.e.}$, Universe stays in false vacuum), keeping the sphalerons in equilibrium, it is pertinent to understand the fate of the false vacuum around $T_c$ and/or $T_n$. Unlike the smooth crossover in the SM where the symmetric phase persists up to 160 GeV, here the Universe can remain trapped in the false vacuum until $T_n$ when broken-phase bubbles begin to envisage the false vacuum space to convert it into the true vacuum inside which sphalerons decouple. To illustrate this, we compute the probability $P(T)$ of a spatial point remaining in the false vacuum at temperature $T$, expressed as~\cite{Guth:1981uk,Guth:1979bh}
\begin{align}
	P(T) & =  e^{-I(T)}, \nonumber \\
	    I(T) & =  \frac{4\pi}{3} \int_T^{T_c} dT' \frac{\Gamma(T')}{T'^4{\mathcal{H}}(T')} \left( \int_T^{T'} d\tilde{T} \frac{v_w}{{\mathcal{H}}(\tilde{T})} \right)^3,
\end{align}
where $v_w (<c_s, ~{\rm {the ~sound ~speed}})$ denotes the bubble wall velocity\footnote{ Determining the bubble wall velocity requires solving the coupled scalar field and plasma transport equations with hydrodynamic matching~\cite{Ai:2023see,Ai:2024btx}, which is beyond the scope of this work. We therefore treat $v_w$ as a phenomenological parameter and assume a subsonic value~\cite{Carena:2022qpf}. Since the lepton asymmetry is generated before bubble nucleation in our framework, the baryon asymmetry is essentially independent of $v_w$, which mainly impacts the gravitational wave prediction though.}. The temperature dependence of $P(T)$ is shown in the lower panel of Fig.~\ref{fig:bubble-number-plot} for $\Lambda =586$ GeV using the potential in Eq.~\ref{high-T-potential}. The result $P(T_n) \simeq 1$ clearly demonstrates that the Universe continues to exist in the false vacuum down to the nucleation temperature $T_n$. In particular, for our case, the Universe remains in false-vacuum down to the lowest temperature $T_n^{\rm min} \simeq 34$ GeV (Eq.~\ref{eq:Lambda-limit}), keeping the sphalerons active and in equilibrium even at $\mathcal{O}(34~\rm GeV)$.

The emergence of such a low sphaleron decoupling temperature in the context of FOEWPT, compared to $T_{\rm sp}^{\rm SM}$, while the Universe stays in the false minima is the key observation of our study, which has a far reaching implications for low scale leptogenesis. Firstly, the Universe being in false vacuum till $T_n$ ($< T_{\rm sp}^{\rm SM}$), the SM fields are massless (apart from their thermal masses), and hence the RHNs of mass $M_N > T_n$ can decay out of equilibrium to lepton and Higgs doublets (via neutrino Yukawa interaction). This requires the satisfaction of the condition, $M_N > M_L(T) + M_H(T)$ where $M_L(T)$ and $M_H(T)$ represent the thermal masses of lepton and Higgs doublets respectively, with $M_L(T) + M_H(T) \simeq 0.77\,T$. Hence, such decay would take place close to a RHN mass-equivalent temperature $T \sim M_N$ which automatically satisfies $M_N > 0.77 \,T$. The exact temperature (or epoch) of occurrence of such RHN decay needs to be estimated by solving the Boltzmann equations as we have established later in this work.
Secondly, the sphalerons are able to convert the lepton asymmetry to baryon one till a temperature at or above $T_n$. For example, $T_n^{\rm min}$ being 34 GeV in our scenario, the standard (resonant) leptogenesis can easily take place via the out-of-equilibrium decay of (quasi-degenerate) RHNs having mass above $T_n^{\rm min}$ but below the SM gauge boson's masses $m_{W,Z}$ ($T_n^{\rm min} < M_N < m_{W,Z}$) in the temperature window: $T_n^{\rm min} - M_N$, which can explain BAU.
Note that producing the correct baryon asymmetry through RHNs out-of-equilibrium decay at such low scale would otherwise remain highly challenging for $M_N < T_{\rm sp}^{\rm SM} \sim 131.7$ GeV even with resonant leptogenesis as conversion of lepton to baryon asymmetry is effectively switched off below $T_{\rm sp}^{\rm SM}$ in the regime of smooth crossover EWPT. Interestingly, our framework is still capable of generating baryon asymmetry in case the Universe attains a low reheating temperature\footnote{In case of non-instantaneous reheating~\cite{Datta:2022jic,Datta:2023pav}, the maximum temperature $T_{\rm Max}$ being more than $T_{\rm RH}$, we expect the present scenario would work with $T_{\rm RH}$ close to the BBN bound.}  $T_{\rm RH}$ below the $T_{\rm sp}^{\rm SM}$, quite plausible as the lower bound on reheating temperature is only a few MeV from BBN ~\cite{Kawasaki:1999na,Kawasaki:2000en,Giudice:2000dp,Giudice:2000ex}, contrary to other low scale leptogenesis scenarios such as Higgs-decay leptogenesis~\cite{Hambye:2016sby}, leptogenesis via oscillations ~\cite{Akhmedov:1998qx} and~\cite{Bhandari:2023wit} where $T_{\rm RH}$ requires to be higher than $T_{\rm sp}^{\rm SM}$. 
It is worth noting that, in our framework, the nucleation temperature $T_n$ sets a lower bound on the RHN mass $M_N$ to reproduce the observed baryon asymmetry, while the cut-off scale $\Lambda$, which also affects $T_n$, imposes an upper bound on $M_N$ from the requirement that the effective framework (responsible for rendering the electroweak phase transition strongly first order via dimension-6 Higgs operator) remains valid.


To proceed estimating the lepton asymmetry with such a low sphaleron decoupling temperature, we begin with the usual Type-I seesaw Lagrangian\footnote{The non-renormalizable $explicit$ lepton-number breaking operator $c\hspace{0.05 cm}\ell_L \ell_L HH/\Lambda_{\rm L}$ can in principle be also present. However, considering only the $soft$ breaking of the lepton number as present in the Majorana mass of the RHNs in Type-I seesaw, effect of this term on neutrino mass can be ignored by considering the coefficient $c$ to be small enough which can be justified with a UV complete picture that is beyond the present discussion.} (in the charged lepton and RHN mass diagonal bases),
\begin{align}
	-\mathcal{L}_{\rm{I}}= \bar{\ell}_{L_\alpha} (Y_\nu)_{\alpha i} \tilde{H} N_i +\frac12 \overline{N_i^c} M_i N_i +h.c.,
	\label{type-I}
\end{align}
where $i=1,2$ (for minimal scenario) and $\alpha=e,~\mu,~\tau$ in general. With two quasi-degenerate RHNs, one can 
evaluate the CP asymmetry, 
\begin{align}
	\varepsilon_{\ell}^i=\sum_{j\neq i}\frac{\text{Im}(Y_\nu^\dagger Y_\nu)_{ij}^2}{(Y_\nu^\dagger Y_\nu)_{ii}(Y_\nu^\dagger Y_\nu)_{jj}} \frac{\left[M_i^2 -M_j^2 \right] M_i \Gamma_{N_j}}{\left[M_i^2 - M_j^2\right]^2+M_i^2\Gamma_{N_j}^2},
\end{align} 
which can be $\sim \mathcal{O}(1)$ if the resonance condition $\Delta M=M_2-M_1 \sim \Gamma_{N_1}/2$ is satisfied. Here $\Gamma_{N_1}$ is the decay rate of $N_1$ to lepton and Higgs doublets. Using Casas-Ibarra (CI) parametrization \cite{Casas:2001sr}, the $Y_{\nu}$ matrix can be constructed using:
	$Y_{\nu}=-i \frac{\sqrt{2}}{v_0} U D_{\sqrt{m}} \mathbf{R} D_{\sqrt{M}}$
where $U$ is the Pontecorvo-Maki-Nakagawa-Sakata matrix~\cite{Zyla:2020zbs} which connects the flavor basis to the mass basis of light neutrinos.  Here $ D_{\sqrt{m}}={\rm{diag}}(\sqrt{m_1},\sqrt{m_2},\sqrt{m_3})$ and $ D_{\sqrt{M}}=\rm{diag}(\sqrt{M_1},\sqrt{M_2})$ denote the diagonal matrices containing the square root of light neutrino masses and RHN masses respectively and $\mathbf{R}(\theta_R)$ represents a complex orthogonal matrix.

\begin{figure}[!htb]
\includegraphics[width=0.94\linewidth]{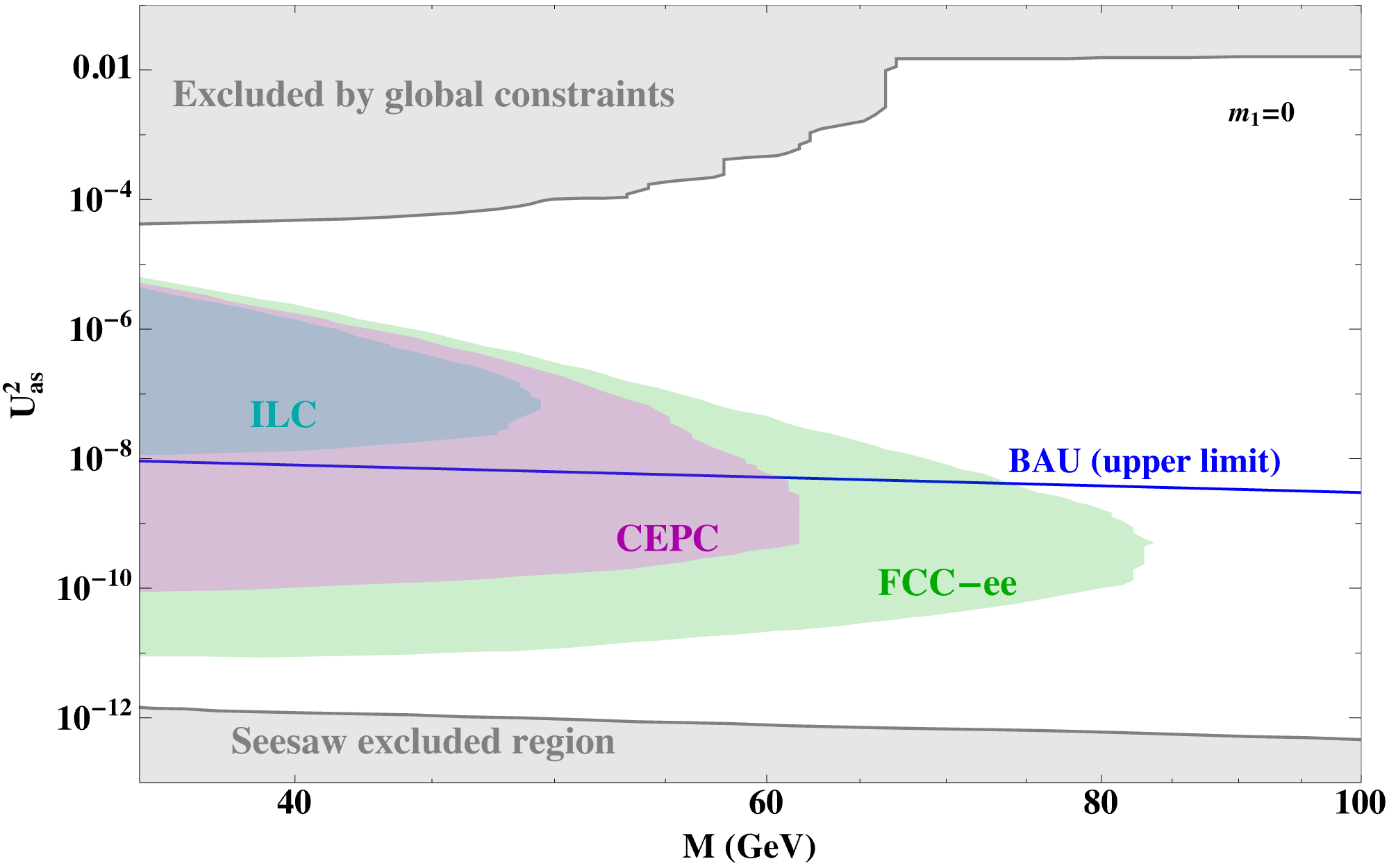}
\caption{Viable parameter space for leptogenesis and sensitivity regions of future lepton colliders~\cite{Antusch:2017pkq} with the upper grey region excluded by global constraints~\cite{Drewes:2016jae} from direct and indirect search experiments.}
\label{fig:M1-Umu2-1}
\end{figure}

The Boltzmann equations for the abundance of RHNs ($Y_N = n_N/s$) and the yield of B-L asymmetry ($Y_{B-L}$) can be written as~\cite{Davidson:2008bu,Giudice:2003jh,Pramanick:2024gvu,Luty:1992un},

\begin{widetext}
\begin{align}
	s {\mathcal{H}} z \frac{d Y_{N_{i}}}{d z}  & = - \left(\frac{Y_{N_{i}}}{Y^{eq}_{N_{i}}} - 1 \right ) \left(\gamma_D^i + 2 \gamma_S^{s} + 4 \gamma_S^{t} \right) ,\\
	s {\mathcal{H}} z \frac{d Y_{B-L}}{d z} & = \sum_{i=1}^{2} \left[ - \varepsilon_{\ell}^i\left(\frac{Y_{N_{i}}}{Y^{eq}_{N_{i}}} - 1 \right)  \gamma_D^i - \frac{Y_{B-L}}{Y_l^{eq}} \left( 2 \gamma_S^{t} + \gamma_S^{s} \frac{Y_{N_i}}{Y_{N_i}^{eq}}  + 2 \gamma_N \right) \right] , 
\end{align}
\end{widetext}
where, $z=M_1/T$, $s$ is the entropy density, and $\gamma_D^i =n_{N_{i}}^{eq} \frac{K_1(M_i/T)}{K_2(M_i/T)} \Gamma_{N_{i}}$. 
Here, $\gamma_S^{s}$ and $\gamma_S^{t}$ denote the reaction rate densities for $\Delta L=1$ scatterings ($s$ and $t$ channels respectively) while $\gamma_N$ represents the reaction rate density of for 
$\Delta L=2$ scattering processes~\cite{Davidson:2008bu,Giudice:2003jh,Pramanick:2024gvu,Luty:1992un} (see appendix~\ref{ap:3} for detailed discussion). Together with decays and inverse decays, these processes account for the relevant washout effects. Considering $T_n (\Lambda) \lesssim M_i < \Lambda$ (as discussed above) while using best fit values of the neutrino mixing angles and mass-square differences with $m_1 = 0$ for normal hierarchy for getting $Y_{\nu}$, we solve the above equations numerically 
with thermalised RHNs as the initial condition. It is found that the correct BAU corresponding to  $M_1 \in [35, 100]$ GeV can be obtained by varying $\Delta M \in \left[3.16 \times 10^{-11}, 1.07 \times 10^{-7}\right]$ and ${\rm{Im}}(\theta_R) \in [-4,-0.2]$ with 
${\rm Re}({\theta_R})$ = 0.8. Correspondingly, the range of neutrino Yukawa coupling (specifically, the largest entry of $\left|Y_\nu \right|$) is found to be $[9.41 \times 10^{-8}, 8.75 \times 10^{-6}]$ while the RHN mass varies within the above specified range.
The ${\rm{Im}}(\theta_R)$ dependence can be conventionally parametrised by the active-sterile mixing angle, $\left|\Theta_{\alpha i}\right|^2 = |(Y_{\nu})_{\alpha i}|^2 v_0^2/M_i^2$. 
The result is depicted in Fig.~\ref{fig:M1-Umu2-1} in the $M$ - $U_{\rm as}^2$ plane, with $M = (M_1 + M_2)/2$ and $U_{\rm as}^2 = \Sigma_{i,\alpha} |\Theta_{\alpha i}|^2$, where the blue solid line corresponds to the upper limit of 
$U^2_{\rm as}$ for correct BAU (as well as neutrino oscillation data) while the region below it stands for more than required BAU, which can, however, be brought down to correct asymmetry easily by appropriate $\Delta M$ and $\theta_R$. 

Note that inclusion of ${(H^{\dagger}H)^3}/{\Lambda^2}$ in the SM Higgs potential modifies the triple Higgs coupling as
\begin{equation}
		\lambda_3 = \frac{1}{6}\left.\frac{d^3 V_{\rm eff}(\phi,T=0)}{d\phi^3}\right |_{\phi=v_0} =  \lambda_3^{\rm SM} +  \frac{v_0^3}{\Lambda^2} ,
\end{equation}
where $\lambda_3^{\rm SM} =  m_h^2/(2 v_0)$. 
\begin{figure}[!htb]
	\includegraphics[width=0.94\linewidth]{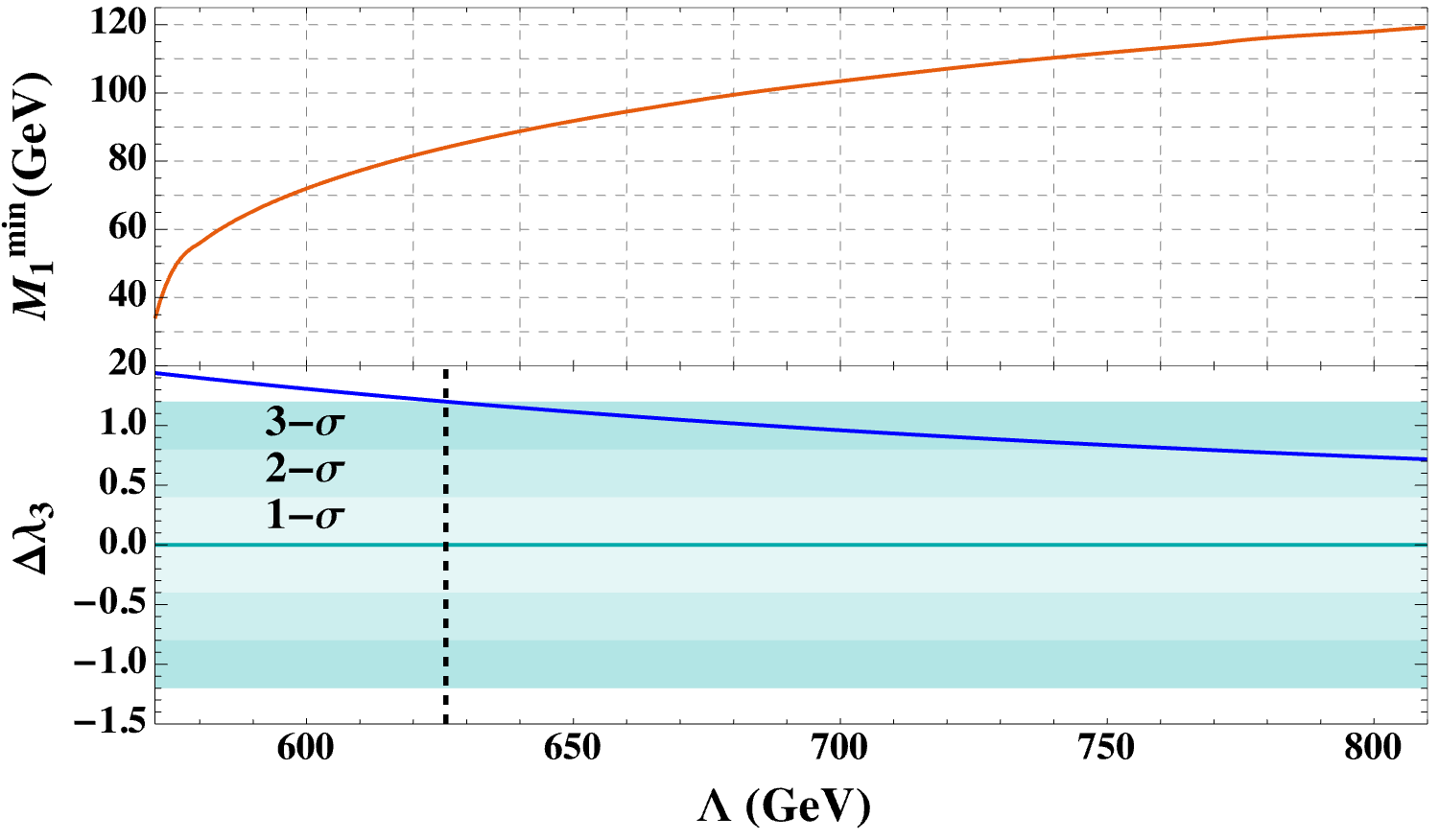}
	\caption{Variation of $\Delta \lambda_3$ (bottom panel) and $M_1^{\rm min}$ (upper panel) with $\Lambda$.}
	\label{fig:Lambda-deltaLm-M1}
\end{figure}
Current bounds from the ATLAS and CMS experiments at 95\% C.L., expressed in terms of $k_3=\lambda_3/\lambda_3^{\rm SM}$ are given by 
\begin{align}
\nonumber
	k_3 \in & \, [-1.2,7.2] \quad {\rm ATLAS}~\text{\cite{ATLAS:2024ish}}, \\
	\nonumber
	k_3 \in & \, [-1.2,7.5] \quad {\rm CMS}~\text{\cite{CMS:2024awa}}.
\end{align}
The lowest value of $\Lambda$ used in our analysis is 571.6 GeV which corresponds to $k_3=2.44$, which lies well within the current experimental bounds of $k_3$ as reported by the ATLAS and CMS collaborations. Analogously, the upper limit of $\Lambda$ ($810$ GeV) obtained in our framework results to $k_3 =1.72$ , hence falls within the allowed range.
 However, the HL-LHC is expected to constrain $\lambda_3$ within 40\% of the SM value at 68\% C.L.~\cite{Goertz:2013kp,Barger:2013jfa,Barr:2014sga}, providing an indirect collider probe of $\Lambda$ and leptogenesis.  
We present $\Delta \lambda_3 =(\lambda_3 - \lambda_3^{\rm SM})/ \lambda_3^{\rm SM}$ as a function of the cutoff scale $\Lambda$, in the range of our interest from FOEWPT and leptogenesis, as shown in Fig.~\ref{fig:Lambda-deltaLm-M1} (bottom panel), with the HL-LHC experimental reach at $1\sigma$, $2\sigma$, and $3\sigma$ indicated by the colored shaded regions. Since the value of $\Lambda$ is intricately connected to the $T_n$ playing significant role in realizing low scale leptogenesis with a minimum mass of RHN $M_1^{\rm min}$ as shown in the upper panel of Fig.~\ref{fig:Lambda-deltaLm-M1}, this serves as an interesting future probe of the low scale leptogenesis, particularly for $\Lambda \gtrsim 625$ GeV (concluded from the vertical dashed line) or for RHN mass $\gtrsim \mathcal{O}(84)$ GeV, 
within $3\sigma$ detection prospects of $\lambda_3$. 
Although the cutoff scale $\Lambda$ in our framework is relatively low, of order ${\cal O}$(600–800) GeV, this does not conflict with current collider constraints, as the new states (for the UV completion of the dimension-6 operator involved) can naturally be heavier than $\Lambda$ while remaining fully consistent with existing searches.

\begin{figure}[!htb]
\includegraphics[width=1\linewidth]{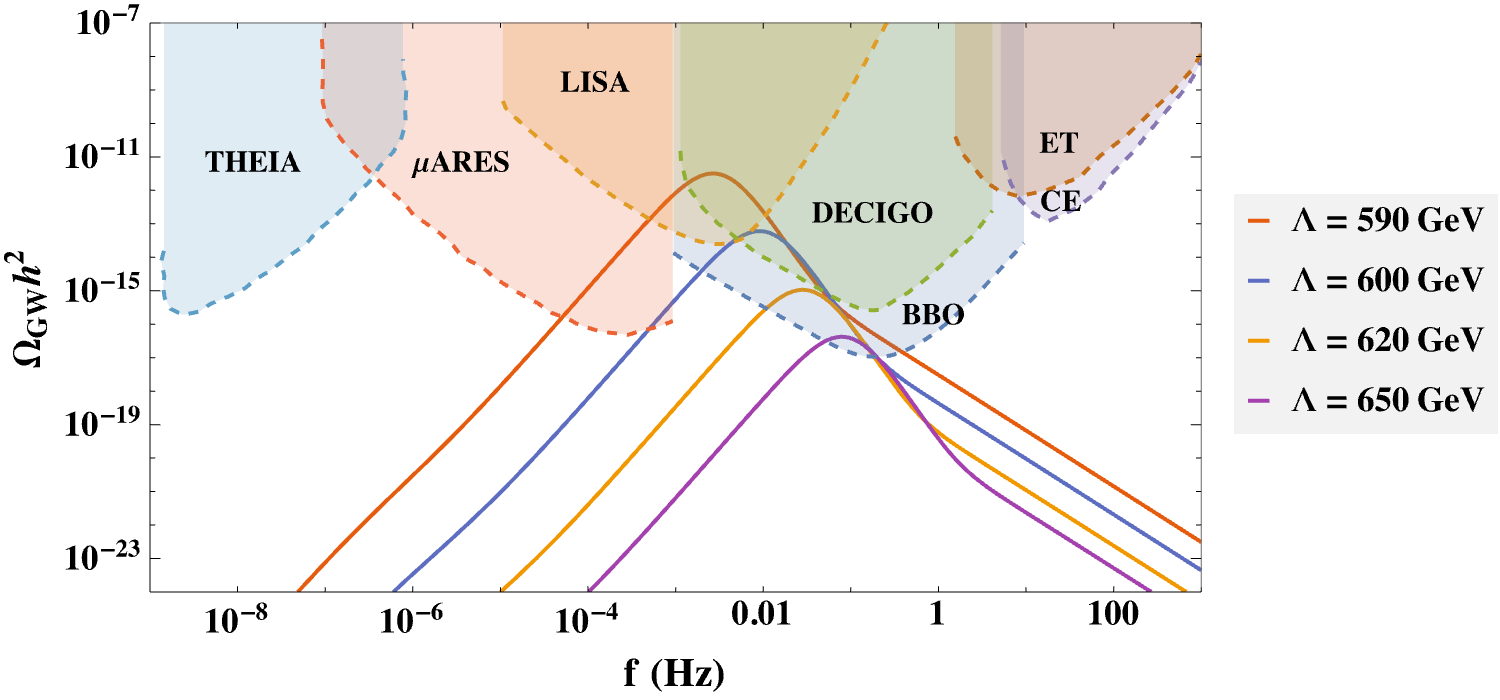}
\caption{Gravitational wave signals with sensitivity ranges for various proposed GW detectors.}
\label{fig: GW plot}
\end{figure}

Furthermore, stochastic gravitational waves (GWs) produced during the SFOEWPT could also be detectable in future GW detectors \cite{Caprini:2015zlo} primarily sourced by the sound waves in the plasma ($\Omega_{\rm sw} h^2$), and magnetohydrodynamic turbulence ($\Omega_{\rm turb} h^2$) in our case. The GW signals with different 
$\Lambda$ values are included in Fig. \ref{fig: GW plot} along with sensitivity regions of various proposed GW detectors 
like LISA \cite{LISA:2017pwj}, BBO \cite{Yagi:2011wg,Crowder:2005nr,Corbin:2005ny,Harry:2006fi}, DECIGO \cite{Yagi:2011wg,Kawamura:2006up}, $\mu$ARES \cite{Sesana:2019vho}, CE \cite{LIGOScientific:2016wof,Reitze:2019iox}, ET \cite{Punturo:2010zz,Hild:2010id,Sathyaprakash:2012jk,ET:2019dnz} and THEIA \cite{Garcia-Bellido:2021zgu}. The involvement of $T_n$ in estimating such signals are elaborated in appendix~\ref{ap:4}. 

To sum up, we find that the sphaleron decoupling temperature can effectively be lowered compared to its SM value $T^{\rm SM}_{\rm sp} = 131.7$ GeV in the context of first order EWPT. This originates due to the existence of unsuppressed sphaleron transition rate till a relatively low bubble nucleation temperature $T_n < T^{\rm SM}_{\rm sp}$, characteristic of the new physics scale (involved in the dimension-6 operator in the Higgs potential) responsible for realizing the EWPT of first order. This conclusion, however, continues to hold for other frameworks of FOEWPT too. 
The strongly first-order electroweak phase transition therefore plays a pivotal role in our framework by lowering the effective sphaleron decoupling temperature, thereby allowing successful low-scale leptogenesis with right handed neutrino masses below the electroweak scale, even lighter than the SM Higgs. In particular, such a finding paves the way of registering a low scale (resonant) leptogenesis scenario where a set of two quasi-degenerate RHNs decay out of equilibrium and produce enough lepton asymmetry below $T <$ 131.7 GeV, which can still be converted to baryon asymmetry. This not only enables the seesaw mechanism to be testable at colliders by lowering down the seesaw scale (the RHN mass $M_N$), but also compatible with low reheating temperature of the Universe, $T_{\rm RH}$ below 131.7 GeV. The latter realization features a unique finding since other existing low scale leptogenesis scenarios, such as via oscillation or Higgs decay, require the reheating  temperature of the Universe to be at or above 131.7 GeV. This proposal also carries profound importance in exploring the sub-EW mass RHNs at future lepton colliders like FCC-ee, CEPC, ILC. The one to one correspondence between the new physics scale $\Lambda$ and $T_n$ (and/or the lower limit of RHN mass scale, $M_N \gtrsim$ 35 GeV) exhibits a tantalizing possibility to prove leptogenesis and seesaw mechanism by measuring the triple Higgs coupling at HL-LHC which may also illuminate upon the era of electroweak symmetry breaking in the early Universe with the complimentary information obtained from the detection of associated GWs. 

\begin{acknowledgements}
The work of DB is supported by Council of Scientific \& Industrial Research (CSIR), Govt. of India, under the senior research fellowship scheme. The work of AS is supported by the grants CRG/2021/005080 and MTR/2021/000774 from SERB, Govt. of India. 
\end{acknowledgements}
\appendix

\onecolumngrid

\section{Effective potential and barrier height dependence on cut-off scale $\Lambda$}\label{ap:1}
\subsection{Effective potential construction} \label{ap:1-1}

In order to study the details of first order electroweak phase transition (FOEWPT), we have to employ the Coleman-Weinberg (zero-temperature) contribution~\cite{Coleman:1973jx} as well as thermal correction~\cite{Quiros:1999jp} to the tree-level potential along with the corrections due to ring diagrams at higher loop, called daisy resummation~\cite{Arnold:1992rz,Carrington:1991hz,Espinosa:1992gq} . The daisy correction has been implemented through Parwani resummation method~\cite{Parwani:1991gq} where zero temperature field dependent masses are replaced by the thermal-corrected field dependent masses in the effective potential.
We can write the full effective potential as
\begin{equation}
	V_{\rm eff}(\phi,T)= V_{(\phi)}^{\rm tree} + V_{\rm CW}(\phi,T) + V_{\rm T}(\phi,T) +\delta V(\phi), \label{effective-potential}
\end{equation}
where, $V_{\rm CW}(\phi,T)$ is the Coleman-Weinberg (CW) potential and in ${\rm \overline{MS}}$ renormalization scheme, we can write as~\cite{Coleman:1973jx}:
\begin{equation}
	V_{CW}(\phi,T)=\frac{1}{64\pi^2}\sum_{i=\phi,G,W,Z,\gamma_L,t}n_i m_i^4(\phi,T)\left[\log\left(\frac{m_i^2(\phi,T)}{\mu^2}\right)-c_i\right] , \label{CW-potential}
\end{equation}
and the thermal correction~\cite{Quiros:1999jp} to the tree-level potential can be written as
\begin{equation}
	V_{T}(\phi,T)=\frac{T^4}{2\pi^2} \left[\sum_{i=\phi,G, W,Z,\gamma_L}n_i J_B\left(\frac{m_i^2(\phi,T)}{T^2}\right)-\sum_{i=t} n_i J_F\left(\frac{m_i^2(\phi,T)}{T^2}\right) \right]. \label{thermal-potential-1}
\end{equation}
The thermal functions $J_{B,F}$ are given by
\begin{equation}
	J_{B,F}\left(\frac{m_i^2(\phi,T)}{T^2}\right)=\int_0^{\infty} dx x^2 \log\left[1\mp {\rm exp}\left(-\sqrt{\frac{x^2+m_i^2(\phi,T)}{T^2}}\right)\right], \label{thermal-integral-1}
\end{equation}
where $\mp$ signs are for bosons and fermions respectively. 
We choose the renormalization scale $\mu=M_Z$, the $Z$ boson mass. The constant $c_i$'s take the value as
\begin{equation*}
	c_{\phi,G,t}=3/2 \hspace{0.4 cm} {\rm and} \hspace{0.4 cm} c_{W,Z,\gamma_L}=5/6,
\end{equation*}
and $n_i$'s denote the number of degrees of freedom of the particles:
\begin{equation*}
	n_{\phi}=1,\hspace{0.25 cm} n_{G}=3,\hspace{0.25 cm} n_{W_T}=4,\hspace{0.25 cm} n_{W_L}=n_{Z_T}=2,\hspace{0.25 cm} n_{Z_L}=n_{\gamma_L}=1,\hspace{0.25 cm} n_t=-12 .
\end{equation*}
The field dependent masses denoted by $m_i(\phi)$ is given by:
\begin{align}
	m_{\phi}^2(\phi) & = -\mu_h^2 + 3 \lambda_h \phi^2 +\frac{15}{4} \frac{\phi^4}{\Lambda^2},  \nonumber \\ 
	m_{G}^2(\phi) & = -\mu_h^2 +  \lambda_h \phi^2 +\frac{3}{4} \frac{\phi^4}{\Lambda^2}, \\ 
	m_{W}^2(\phi) & =  \frac{g_{W}^2}{4} \phi^2, \quad 
	m_{Z}^2(\phi) = \frac{g_{W}^2 + g_{Y}^2}{4} \phi^2, \quad  
	m_t^2(\phi)  = \frac{y_t^2}{2} \phi^2,  \nonumber
\end{align}
and the self energy $\Pi_i(T)$ of the corresponding fields take the form~\cite{Quiros:1999jp}:
\begin{align}
	& \Pi_h(T) = \Pi_{G}(T)=\frac{1}{16}\left( \frac{4 m_{h}^2}{v_0^2} + 3 g_{\rm w}^2 + g_Y^2 +4 y_t^2 -12 \frac{v_0^2}{\Lambda^2} \right)T^2, \nonumber \\
	& \Pi_{W_L}(T) = \frac{11}{6}g_{\rm w}^2 T^2, \nonumber \\
	& \Pi_{W_T} = \Pi_{Z_T}=\Pi_{\gamma_T}=0.
\end{align}
The thermal-corrected field dependent mass is defined by  $m_i^2(\phi,T)=m_i^2(\phi)+\Pi_i(T)$. Accordingly, the thermal-corrected masses of the longitudinal mode of $Z$ boson and the photon are
\begin{align}
	& m_{Z_L}^2(\phi,T)=\frac{1}{2}\left[m_{\rm Z}^2(\phi)+\Pi_{W_L}(T)+ \frac{11}{6}g_{\rm y}^2 T^2+\Delta(\phi,T)\right], \\
	& m_{\gamma_L}^2(\phi,T)=\frac{1}{2}\left[m_{\rm Z}^2(\phi)+\Pi_{W_L}(T)+ \frac{11}{6}g_{\rm y}^2 T^2-\Delta(\phi,T)\right],
\end{align}
where
\begin{equation*} \Delta(\phi,T)=\sqrt{(g_{\rm w}^2-g_{\rm y}^2)^2\left(\frac{1}{4}\phi^2+\frac{11}{6}T^2\right)^2+\frac{1}{4}g_{\rm w}^2 g_{\rm y}^2 \phi^4} \quad .
\end{equation*}

\noindent The last term in Eq.~\ref{effective-potential}, accounts for the counterterms, is expressed as:
\begin{equation}
	\delta V(\phi)= A \phi^2+B \phi^4 .
\end{equation}
The coefficients $A$ and $B$ are determined by imposing the renormalization conditions
\begin{align}
	\left.\frac{\partial V_{\rm eff}(\phi,T)}{\partial \phi}\right |_{\phi=v_0,T=0} &=0 , \nonumber \\
	\left.\frac{\partial^2 V_{\rm eff}(\phi,T)}{\partial \phi^2}\right |_{\phi=v_0,T=0} &=m_h^2  .
\end{align}
We then employ the full effective thermal potential, including the Coleman–Weinberg term, the daisy resummation and the counterterms, using the {\bf CosmoTransition} package~\cite{Wainwright:2011kj} to generate the final results of the manuscript. We then extract the values of $\Lambda$ that yield a successful first-order electroweak phase transition and use them to generate the final parameter space described in the main manuscript.

Taking high-temperature expansion and neglecting the CW correction term, we restrict ourselves terms upto order $T^2$ and hence ignoring the daisy corrections also, the effective potential Eq.~\ref{effective-potential} can be written as:
\begin{equation}
	V_{\rm eff}(\phi,T) =  \left( -\frac{\mu_h^2}{2}+\frac{1}{2}c_h T^2 \right)\phi^2 + \left( \frac{\lambda_h}{4} + \frac{\lambda_1}{4}T^2 \right) \phi^4 + \frac{1}{8}\frac{\phi^6}{\Lambda^2}, \label{high-T-potential-1}
\end{equation}
where
\begin{equation}
	c_h=\frac{1}{16}\left( \frac{4 m_{h}^2}{v_0^2} + 3 g_{\rm w}^2 + g_Y^2 +4 y_t^2 -12 \frac{v_0^2}{\Lambda^2} \right), \hspace{0.28 cm} \lambda_1=\frac{1}{\Lambda^2}. \label{ch-lamda1-value-1}
\end{equation}
Here, Eq.~\ref{high-T-potential-1} represents the high-temperature approximated form of the effective Higgs potential considered in the first part of the main manuscript. 
The parameters $\mu_h$ and $\lambda_h$ can be determined using the conditions:
\begin{equation}
	\left. \frac{d V_{\rm eff}(\phi,T=0)}{d \phi} \right|_{\phi=v_0}=0 \hspace{0.1 cm}, \quad 	\left. \frac{d^2 V_{\rm eff}(\phi,T=0)}{d \phi^2} \right|_{\phi=v_0}=m_h^2 \label{condition-eq}.
\end{equation}

\subsection{Barrier height dependence on $\Lambda$}  \label{ap:1-2}
\begin{figure}[!htb]
	\includegraphics[width=0.6\linewidth]{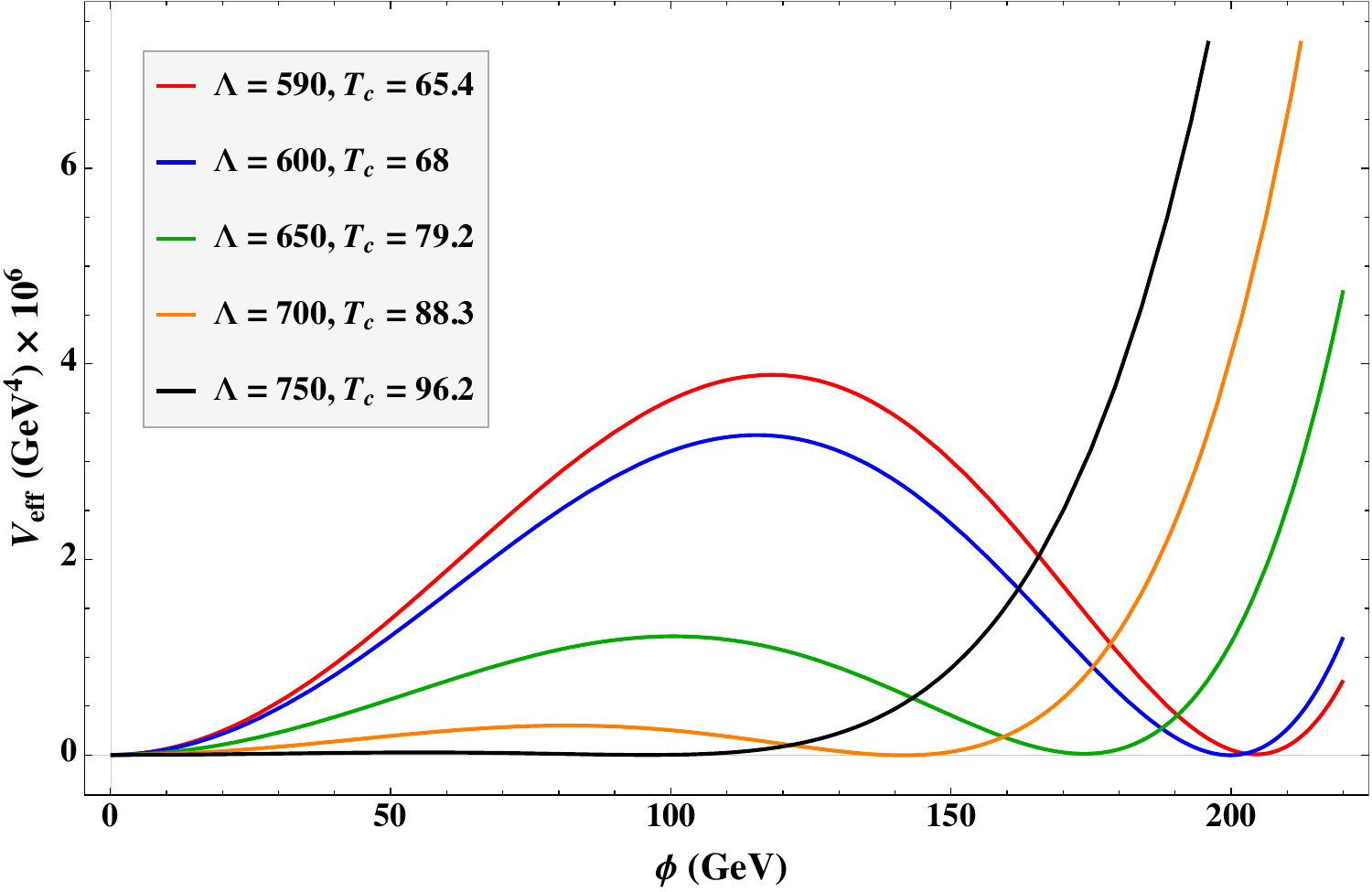}
	\caption{Potential at critical temperature $T_c$ (in GeV) for several values of $\Lambda$ (in GeV).}
	\label{fig:veff-phi-plot}
\end{figure}

Figure~\ref{fig:veff-phi-plot} shows the effective potential, Eq.~\ref{high-T-potential-1}, for different $\Lambda$ values, corresponding to different critical temperatures $T_c$. As $\Lambda$ increases, the barrier height decreases and vice-versa. 
For $\Lambda = 590$ GeV, the barrier height is significant, while at $\Lambda=750$ GeV, it is noticeably reduced. Beyond $\Lambda = 750$ GeV, the barrier becomes too small to take place a first order phase transition, establishing this as the upper limit for a successful transition.

\section{Euclidean action calculation}\label{ap:2}
The tunnelling rate of the universe from the false vacuum to true vacuum phase is given by $\Gamma(T) \simeq T^4 \left( \frac{S_3}{2\pi T} \right)^{3/2} {\rm exp}(-S_3/T)$, where $S_3$ is the three-dimensional Euclidean action for an $O(3)$ symmetric~\cite{Linde:1981zj} bubble configuration. The bubble configuration, also called the bounce solution~\cite{Coleman:1977py}, can be found by solving the equation
\begin{equation}
	\frac{d^2\phi}{d r^2} +  \frac{2}{r}  \frac{d\phi}{d r} -  \frac{\partial V_{\rm eff}(\phi,T)}{\partial \phi}=0, \label{bounce-equation}
\end{equation}
with boundary conditions
\begin{equation}
	\phi(r\rightarrow \infty)=0 \quad {\rm and}  \quad  \frac{d\phi(r=0)}{d r}=0,
\end{equation}
where $r$ is the distance from the center of the bubble. The action $S_3$ can be written as
\begin{equation}
	S_3 = 4\pi \int dr r^2  \left(   \frac{1}{2}  \left( \frac{d\phi}{d r}\right)^2 + V_{\rm eff}(\phi,T)  \right).
\end{equation}
Eq.~(\ref{bounce-equation}) can be solved numerically using the Mathematica Package {\bf FindBounce}~\cite{Guada:2020xnz} and then using the bounce solution we can determine $S_3$ as a function of temperature.

\section{Reaction Densities of $\Delta L=1$ and $\Delta L=2$ scattering processes} \label{ap:3}
We have incorporated the full Boltzmann equations, including all $\Delta L = 1$ and $\Delta L = 2$ scattering processes of RHNs with the SM particles, together with the decay and inverse decay channels. The relevant scattering processes are listed below:
\begin{align}
	\textbf{$\mathbf{\Delta L=1}$ scattering:} \quad
	& \text{i)}\; N_1 + \ell_L \leftrightarrow Q + \overline{U} ,
	&& \text{ii)}\; N_1 + \ell_L \leftrightarrow \overline{H} + V_{\mu}, \nonumber \\
	&  \text{iii)}\; N_1 + \overline{Q} \leftrightarrow \overline{\ell}_L + \overline{U},
	&&  \text{iv)}\; N_1 + U \leftrightarrow \overline{\ell}_L + \overline{Q}, \nonumber \\
	& \text{v)}\; N_1 + V_{\mu} \leftrightarrow \overline{\ell}_L + \overline{H},
	&& \text{vi)}\; N_1 + H \leftrightarrow \overline{\ell}_L + V_{\mu}, \\[6pt]
	\textbf{$\mathbf{\Delta L=2}$ scattering:} \quad
	& \text{i)}\; \ell_L + H \leftrightarrow \overline{\ell}_L + \overline{H},
	&& \text{ii)}\; \ell_L + \ell_L \leftrightarrow \overline{H} + \overline{H}.
\end{align} 
For a general $2 \to 2$ scattering process, the reaction density in the centre of mass frame can be expressed as~\cite{Davidson:2008bu,Giudice:2003jh,Pramanick:2024gvu} 
\begin{align}
	\gamma_{12\leftrightarrow 34} = \frac{T}{64\pi^4} \int_{s_{\rm min}}^{\inf} ds K_1(z\sqrt{s}) \sqrt{s} \hat{\sigma}(s),
\end{align}
where $s$ denotes the centre of mass energy squared. The reduced cross section, defined as $\hat{\sigma}(s)$, is given by~\cite{Luty:1992un}
\begin{equation}
	\hat{\sigma}(s)= \frac{1}{8\pi s} \int_{t_-}^{t_+} dt \left| \mathcal{M}_{12\leftrightarrow 34}(s,t) \right|^2.
\end{equation}
Here, $s_{\rm min} = {\rm Max}[(m_1 + m_2)^2, (m_3 + m_4)^2]$, and $m_i$ $(i=1,2,3,4)$ denote the masses of the particles participating in the scattering. The quantity $\left| \mathcal{M}_{12\leftrightarrow 34} \right|^2$ represents the squared matrix element, while $t$ is the Mandelstam variable with the integration limits are given by~\cite{Pramanick:2024gvu}
\begin{equation}
	t_{\pm} = \frac{1}{4 s} \left[(m_1^2 - m_2^2 - m_3^2 + m_4^2)^2 - \left(\lambda^{1/2}(s,m_1^2,m_2^2) \mp \lambda^{1/2}(s,m_3^2,m_4^2)\right)^2\right].
\end{equation}
$\lambda(p,q,r)$ is known as Kallen function and can be defined as 
\begin{equation}
	\lambda(p,q,r) = p^2 + q^2 + r^2 - 2pq -2pr-2qr.
\end{equation} 
The quantity $\gamma_S^{s}$ includes the $s$-channel $\Delta L=1 $ scattering processes $[N_1 + \ell_L \leftrightarrow Q + \overline{U}]$ and $[N_1 + \ell_L \leftrightarrow \overline{H} + V_{\mu}]$. 
The term $\gamma_S^{t}$ accounts for the $t$-channel $\Delta L=1 $ scattering processes $[N_1 + \overline{Q} \leftrightarrow \overline{\ell}_L + \overline{U}]$, $[N_1 + U \leftrightarrow \overline{\ell}_L + \overline{Q}]$, $[N_1 + V_{\mu} \leftrightarrow \overline{\ell}_L + \overline{H}]$, and $[N_1 + H \leftrightarrow \overline{\ell}_L + V_{\mu}]$. The contribution from the $\Delta L=2$ scattering processes $[ \ell_L + H \leftrightarrow \overline{\ell}_L + \overline{H}]$ and $[ \ell_L + \ell_L \leftrightarrow \overline{H} + \overline{H}]$ is incorporated through the term $\gamma_N$.

\section{Gravitational wave spectrum calculation}\label{ap:4}
The primary sources of GW production in a cosmological first-order phase transition include bubble collisions, sound waves in the plasma, and magnetohydrodynamic (MHD) turbulence. The relative significance of these sources depends on the phase transition dynamics, particularly the bubble wall velocity $v_w$.\\
In this work, we consider the non-runaway bubble wall scenario, where the wall reaches a subsonic terminal velocity $v_w<c_s$ with $c_s=1/\sqrt{3}$ being the sound speed in the plasma.
In this regime, the dominant contributions to the GW spectrum arise from sound waves and MHD turbulence in the plasma.  The total GW spectrum can be expressed as~\cite{Caprini:2015zlo}
\begin{equation}
	\Omega_{\rm GW}h^2 \simeq \Omega_{\rm sw} h^2 + \Omega_{\rm turb} h^2.
\end{equation}
The contributions from the sound waves and MHD turbulence are, respectively, given by~\cite{Athron:2023xlk,Guo:2020grp,Caprini:2019egz,Hindmarsh:2020hop}
\begin{align}
	\Omega_{\rm sw}(f) h^2 & =\frac{1.23 \times 10^{-5}}{g_*^{1/3}} \frac{H_*}{\beta} \left( \frac{k_{\rm sw} \alpha_*}{1+\alpha_*} \right)^2 v_w S_{\rm sw}(f) \Upsilon , \\
	\Omega_{\rm turb}(f) h^2 & =\frac{1.55 \times 10^{-3}}{g_*^{1/3}} \frac{H_*}{\beta} \left( \frac{k_{\rm turb} \alpha_*}{1+\alpha_*} \right)^{\frac{3}{2}} v_w S_{\rm turb}(f).
\end{align}
Here, $\Upsilon = 1 - \frac{1}{\sqrt{1 + 2 \tau_{\rm sw}  H_*}}$ is a suppression factor~\cite{Guo:2020grp} where $\tau_{\rm sw} \sim R_*/\bar{U_f} $ with $R_* = (8\pi)^{1/3} v_w \beta^{-1}$ be the mean bubble separation and $\bar{U_f}= \sqrt{\frac{3 k_{\rm sw} \alpha_*}{4(1 + \alpha_*)}} $ be the rms fluid velocity.\\
The fraction $\beta/H_*$, with $\beta$ being the inverse time duration of the phase transition
and $H_*$ being the Hubble constant at temperature $T_n$, can be evaluated as~\cite{Grojean:2006bp}
\begin{equation}
	\frac{\beta}{H_*} \simeq T_n \left. \frac{d}{dT}\left( \frac{S_3}{T} \right) \right|_{T=T_n}
\end{equation}
assuming $T_n$ is the temperature of the plasma when GW is produced.\\
The quantity $\alpha_*$, denoting the ratio of released vacuum energy in the phase transition to that of the radiation bath at $T=T_n$, can be written as~\cite{Grojean:2006bp}
\begin{equation}
	\alpha_* = \frac{1}{\rho^*_{\rm rad}} \left[ \Delta V - \frac{T}{4} \frac{\partial \Delta V}{\partial T} \right]_{T=T_n},
\end{equation}
where $\Delta V = V_{\rm eff}(\phi_{\rm false},T)-V_{\rm eff}(\phi_{\rm true},T)$ and $\rho^*_{\rm rad} = g_* \pi^2 T_n^4/30$ is the radiation energy density at $T_n$.

The functions parametrizing the spectral shape of the GWs read as~\cite{Caprini:2015zlo,Vaskonen:2016yiu}
\begin{align}
	S_{\rm sw}(f) = \left( \frac{f}{f_{\rm sw}} \right)^3 \left( \frac{7}{4+3(f/f_{\rm sw})^2}  \right)^{7/2} ,\quad
	S_{\rm turb}(f) = \frac{(f/f_{\rm turb})^3}{(1+ (f/f_{\rm trub}))^{\frac{11}{3}} (1+ 8\pi f/h_*)},
\end{align}
with 
\begin{equation}
	h_*=1.65 \times 10^{-5} \hspace{0.04 cm}{\rm Hz} \left( \frac{T_n}{100 \hspace{0.04 cm} {\rm GeV}} \right)  \left( \frac{g_*}{100 } \right)^{\frac{1}{6}}.
\end{equation}
Here $f_{\rm sw}$ and $f_{\rm turb}$ are the peak frequencies of each contribution can be written as
\begin{align}
	f_{\rm sw}  = \frac{1.9 \times 10^{-5} \hspace{0.04 cm}{\rm Hz}}{v_w} \frac{\beta}{H_*} \left( \frac{T_n}{100 \hspace{0.04 cm} {\rm GeV}} \right)  \left( \frac{g_*}{100 } \right)^{\frac{1}{6}} , \quad
	f_{\rm turb} = 1.42 f_{\rm sw} .
\end{align}
$k_{\rm sw}$ and $k_{\rm turb}$ are the fractions of the released vacuum energy density converted into bulk motion of fluid and MHD turbulence, respectively. For subsonic bubble walls these can be defined as~\cite{Espinosa:2010hh}
\begin{align}
	k_{\rm sw}  = \frac{c_s^{11/5} k_{\rm a}k_{\rm b}}{(c_s^{11/5}-v_w^{11/5})k_{\rm b} + v_w c_s^{6/5} k_{\rm a}} , \quad
	k_{\rm turb}  = \epsilon k_{\rm sw} ,
\end{align}
with $\epsilon$ typically in the range of $5\%$-$10\%$~\cite{Caprini:2015zlo,Hindmarsh:2015qta}. We use $\epsilon=0.05$ for conservative choice. $k_{\rm a}$ and $k_{\rm b}$ are defined as~\cite{Espinosa:2010hh}
\begin{align}
	k_{\rm a} = \frac{6.9 v_w^{6/5}\alpha_*}{1.36 - 0.037 \sqrt{\alpha_*} + \alpha_*} ,\quad
	k_{\rm b} = \frac{\alpha_*^{2/5}}{0.017 + (0.997 + \alpha_*)^{2/5}}.
\end{align}

\twocolumngrid
\bibliography{Ref-3rdwork.bib}

\begin{thebibliography}{96}%
\makeatletter
\providecommand \@ifxundefined [1]{%
 \@ifx{#1\undefined}
}%
\providecommand \@ifnum [1]{%
 \ifnum #1\expandafter \@firstoftwo
 \else \expandafter \@secondoftwo
 \fi
}%
\providecommand \@ifx [1]{%
 \ifx #1\expandafter \@firstoftwo
 \else \expandafter \@secondoftwo
 \fi
}%
\providecommand \natexlab [1]{#1}%
\providecommand \enquote  [1]{``#1''}%
\providecommand \bibnamefont  [1]{#1}%
\providecommand \bibfnamefont [1]{#1}%
\providecommand \citenamefont [1]{#1}%
\providecommand \href@noop [0]{\@secondoftwo}%
\providecommand \href [0]{\begingroup \@sanitize@url \@href}%
\providecommand \@href[1]{\@@startlink{#1}\@@href}%
\providecommand \@@href[1]{\endgroup#1\@@endlink}%
\providecommand \@sanitize@url [0]{\catcode `\\12\catcode `\$12\catcode
  `\&12\catcode `\#12\catcode `\^12\catcode `\_12\catcode `\%12\relax}%
\providecommand \@@startlink[1]{}%
\providecommand \@@endlink[0]{}%
\providecommand \url  [0]{\begingroup\@sanitize@url \@url }%
\providecommand \@url [1]{\endgroup\@href {#1}{\urlprefix }}%
\providecommand \urlprefix  [0]{URL }%
\providecommand \Eprint [0]{\href }%
\providecommand \doibase [0]{https://doi.org/}%
\providecommand \selectlanguage [0]{\@gobble}%
\providecommand \bibinfo  [0]{\@secondoftwo}%
\providecommand \bibfield  [0]{\@secondoftwo}%
\providecommand \translation [1]{[#1]}%
\providecommand \BibitemOpen [0]{}%
\providecommand \bibitemStop [0]{}%
\providecommand \bibitemNoStop [0]{.\EOS\space}%
\providecommand \EOS [0]{\spacefactor3000\relax}%
\providecommand \BibitemShut  [1]{\csname bibitem#1\endcsname}%
\let\auto@bib@innerbib\@empty
\bibitem [{\citenamefont {Aghanim}\ \emph {et~al.}(2020)\citenamefont {Aghanim}
  \emph {et~al.}}]{Planck:2018vyg}%
  \BibitemOpen
  \bibfield  {author} {\bibinfo {author} {\bibfnamefont {N.}~\bibnamefont
  {Aghanim}} \emph {et~al.} (\bibinfo {collaboration} {Planck}),\ }\bibfield
  {title} {\bibinfo {title} {{Planck 2018 results. VI. Cosmological
  parameters}},\ }\href {https://doi.org/10.1051/0004-6361/201833910}
  {\bibfield  {journal} {\bibinfo  {journal} {Astron. Astrophys.}\ }\textbf
  {\bibinfo {volume} {641}},\ \bibinfo {pages} {A6} (\bibinfo {year} {2020})},\
  \bibinfo {note} {[Erratum: Astron.Astrophys. 652, C4 (2021)]},\ \Eprint
  {https://arxiv.org/abs/1807.06209} {arXiv:1807.06209 [astro-ph.CO]}
  \BibitemShut {NoStop}%
\bibitem [{\citenamefont {Luty}(1992)}]{Luty:1992un}%
  \BibitemOpen
  \bibfield  {author} {\bibinfo {author} {\bibfnamefont {M.~A.}\ \bibnamefont
  {Luty}},\ }\bibfield  {title} {\bibinfo {title} {{Baryogenesis via
  leptogenesis}},\ }\href {https://doi.org/10.1103/PhysRevD.45.455} {\bibfield
  {journal} {\bibinfo  {journal} {Phys. Rev. D}\ }\textbf {\bibinfo {volume}
  {45}},\ \bibinfo {pages} {455} (\bibinfo {year} {1992})}\BibitemShut
  {NoStop}%
\bibitem [{\citenamefont {Fukugita}\ and\ \citenamefont
  {Yanagida}(1986)}]{Fukugita:1986hr}%
  \BibitemOpen
  \bibfield  {author} {\bibinfo {author} {\bibfnamefont {M.}~\bibnamefont
  {Fukugita}}\ and\ \bibinfo {author} {\bibfnamefont {T.}~\bibnamefont
  {Yanagida}},\ }\bibfield  {title} {\bibinfo {title} {{Baryogenesis Without
  Grand Unification}},\ }\href {https://doi.org/10.1016/0370-2693(86)91126-3}
  {\bibfield  {journal} {\bibinfo  {journal} {Phys. Lett. B}\ }\textbf
  {\bibinfo {volume} {174}},\ \bibinfo {pages} {45} (\bibinfo {year}
  {1986})}\BibitemShut {NoStop}%
\bibitem [{\citenamefont {Plumacher}(1997)}]{Plumacher:1996kc}%
  \BibitemOpen
  \bibfield  {author} {\bibinfo {author} {\bibfnamefont {M.}~\bibnamefont
  {Plumacher}},\ }\bibfield  {title} {\bibinfo {title} {{Baryogenesis and
  lepton number violation}},\ }\href {https://doi.org/10.1007/s002880050418}
  {\bibfield  {journal} {\bibinfo  {journal} {Z. Phys. C}\ }\textbf {\bibinfo
  {volume} {74}},\ \bibinfo {pages} {549} (\bibinfo {year} {1997})},\ \Eprint
  {https://arxiv.org/abs/hep-ph/9604229} {arXiv:hep-ph/9604229} \BibitemShut
  {NoStop}%
\bibitem [{\citenamefont {Covi}\ \emph {et~al.}(1996)\citenamefont {Covi},
  \citenamefont {Roulet},\ and\ \citenamefont {Vissani}}]{Covi:1996wh}%
  \BibitemOpen
  \bibfield  {author} {\bibinfo {author} {\bibfnamefont {L.}~\bibnamefont
  {Covi}}, \bibinfo {author} {\bibfnamefont {E.}~\bibnamefont {Roulet}},\ and\
  \bibinfo {author} {\bibfnamefont {F.}~\bibnamefont {Vissani}},\ }\bibfield
  {title} {\bibinfo {title} {{CP violating decays in leptogenesis scenarios}},\
  }\href {https://doi.org/10.1016/0370-2693(96)00817-9} {\bibfield  {journal}
  {\bibinfo  {journal} {Phys. Lett. B}\ }\textbf {\bibinfo {volume} {384}},\
  \bibinfo {pages} {169} (\bibinfo {year} {1996})},\ \Eprint
  {https://arxiv.org/abs/hep-ph/9605319} {arXiv:hep-ph/9605319} \BibitemShut
  {NoStop}%
\bibitem [{\citenamefont {Kuzmin}\ \emph {et~al.}(1985)\citenamefont {Kuzmin},
  \citenamefont {Rubakov},\ and\ \citenamefont {Shaposhnikov}}]{Kuzmin:1985mm}%
  \BibitemOpen
  \bibfield  {author} {\bibinfo {author} {\bibfnamefont {V.~A.}\ \bibnamefont
  {Kuzmin}}, \bibinfo {author} {\bibfnamefont {V.~A.}\ \bibnamefont
  {Rubakov}},\ and\ \bibinfo {author} {\bibfnamefont {M.~E.}\ \bibnamefont
  {Shaposhnikov}},\ }\bibfield  {title} {\bibinfo {title} {{On the Anomalous
  Electroweak Baryon Number Nonconservation in the Early Universe}},\ }\href
  {https://doi.org/10.1016/0370-2693(85)91028-7} {\bibfield  {journal}
  {\bibinfo  {journal} {Phys. Lett. B}\ }\textbf {\bibinfo {volume} {155}},\
  \bibinfo {pages} {36} (\bibinfo {year} {1985})}\BibitemShut {NoStop}%
\bibitem [{\citenamefont {Arnold}\ and\ \citenamefont
  {McLerran}(1987)}]{Arnold:1987mh}%
  \BibitemOpen
  \bibfield  {author} {\bibinfo {author} {\bibfnamefont {P.~B.}\ \bibnamefont
  {Arnold}}\ and\ \bibinfo {author} {\bibfnamefont {L.~D.}\ \bibnamefont
  {McLerran}},\ }\bibfield  {title} {\bibinfo {title} {{Sphalerons, Small
  Fluctuations and Baryon Number Violation in Electroweak Theory}},\ }\href
  {https://doi.org/10.1103/PhysRevD.36.581} {\bibfield  {journal} {\bibinfo
  {journal} {Phys. Rev. D}\ }\textbf {\bibinfo {volume} {36}},\ \bibinfo
  {pages} {581} (\bibinfo {year} {1987})}\BibitemShut {NoStop}%
\bibitem [{\citenamefont {Arnold}\ and\ \citenamefont
  {McLerran}(1988)}]{Arnold:1987zg}%
  \BibitemOpen
  \bibfield  {author} {\bibinfo {author} {\bibfnamefont {P.~B.}\ \bibnamefont
  {Arnold}}\ and\ \bibinfo {author} {\bibfnamefont {L.~D.}\ \bibnamefont
  {McLerran}},\ }\bibfield  {title} {\bibinfo {title} {{The Sphaleron Strikes
  Back}},\ }\href {https://doi.org/10.1103/PhysRevD.37.1020} {\bibfield
  {journal} {\bibinfo  {journal} {Phys. Rev. D}\ }\textbf {\bibinfo {volume}
  {37}},\ \bibinfo {pages} {1020} (\bibinfo {year} {1988})}\BibitemShut
  {NoStop}%
\bibitem [{\citenamefont {Bochkarev}\ and\ \citenamefont
  {Shaposhnikov}(1987)}]{Bochkarev:1987wf}%
  \BibitemOpen
  \bibfield  {author} {\bibinfo {author} {\bibfnamefont {A.~I.}\ \bibnamefont
  {Bochkarev}}\ and\ \bibinfo {author} {\bibfnamefont {M.~E.}\ \bibnamefont
  {Shaposhnikov}},\ }\bibfield  {title} {\bibinfo {title} {{Electroweak
  Production of Baryon Asymmetry and Upper Bounds on the Higgs and Top
  Masses}},\ }\href {https://doi.org/10.1142/S0217732387000537} {\bibfield
  {journal} {\bibinfo  {journal} {Mod. Phys. Lett. A}\ }\textbf {\bibinfo
  {volume} {2}},\ \bibinfo {pages} {417} (\bibinfo {year} {1987})}\BibitemShut
  {NoStop}%
\bibitem [{\citenamefont {Khlebnikov}\ and\ \citenamefont
  {Shaposhnikov}(1988)}]{Khlebnikov:1988sr}%
  \BibitemOpen
  \bibfield  {author} {\bibinfo {author} {\bibfnamefont {S.~Y.}\ \bibnamefont
  {Khlebnikov}}\ and\ \bibinfo {author} {\bibfnamefont {M.~E.}\ \bibnamefont
  {Shaposhnikov}},\ }\bibfield  {title} {\bibinfo {title} {{The Statistical
  Theory of Anomalous Fermion Number Nonconservation}},\ }\href
  {https://doi.org/10.1016/0550-3213(88)90133-2} {\bibfield  {journal}
  {\bibinfo  {journal} {Nucl. Phys. B}\ }\textbf {\bibinfo {volume} {308}},\
  \bibinfo {pages} {885} (\bibinfo {year} {1988})}\BibitemShut {NoStop}%
\bibitem [{\citenamefont {Minkowski}(1977)}]{Minkowski:1977sc}%
  \BibitemOpen
  \bibfield  {author} {\bibinfo {author} {\bibfnamefont {P.}~\bibnamefont
  {Minkowski}},\ }\bibfield  {title} {\bibinfo {title} {{$\mu \to e\gamma$ at a
  Rate of One Out of $10^{9}$ Muon Decays?}},\ }\href
  {https://doi.org/10.1016/0370-2693(77)90435-X} {\bibfield  {journal}
  {\bibinfo  {journal} {Phys. Lett. B}\ }\textbf {\bibinfo {volume} {67}},\
  \bibinfo {pages} {421} (\bibinfo {year} {1977})}\BibitemShut {NoStop}%
\bibitem [{\citenamefont {Gell-Mann}\ \emph {et~al.}(1979)\citenamefont
  {Gell-Mann}, \citenamefont {Ramond},\ and\ \citenamefont
  {Slansky}}]{Gell-Mann:1979vob}%
  \BibitemOpen
  \bibfield  {author} {\bibinfo {author} {\bibfnamefont {M.}~\bibnamefont
  {Gell-Mann}}, \bibinfo {author} {\bibfnamefont {P.}~\bibnamefont {Ramond}},\
  and\ \bibinfo {author} {\bibfnamefont {R.}~\bibnamefont {Slansky}},\
  }\bibfield  {title} {\bibinfo {title} {{Complex Spinors and Unified
  Theories}},\ }\href@noop {} {\bibfield  {journal} {\bibinfo  {journal} {Conf.
  Proc. C}\ }\textbf {\bibinfo {volume} {790927}},\ \bibinfo {pages} {315}
  (\bibinfo {year} {1979})},\ \Eprint {https://arxiv.org/abs/1306.4669}
  {arXiv:1306.4669 [hep-th]} \BibitemShut {NoStop}%
\bibitem [{\citenamefont {Mohapatra}\ and\ \citenamefont
  {Senjanovic}(1980)}]{Mohapatra:1979ia}%
  \BibitemOpen
  \bibfield  {author} {\bibinfo {author} {\bibfnamefont {R.~N.}\ \bibnamefont
  {Mohapatra}}\ and\ \bibinfo {author} {\bibfnamefont {G.}~\bibnamefont
  {Senjanovic}},\ }\bibfield  {title} {\bibinfo {title} {{Neutrino Mass and
  Spontaneous Parity Nonconservation}},\ }\href
  {https://doi.org/10.1103/PhysRevLett.44.912} {\bibfield  {journal} {\bibinfo
  {journal} {Phys. Rev. Lett.}\ }\textbf {\bibinfo {volume} {44}},\ \bibinfo
  {pages} {912} (\bibinfo {year} {1980})}\BibitemShut {NoStop}%
\bibitem [{\citenamefont {Yanagida}(1979)}]{Yanagida:1979as}%
  \BibitemOpen
  \bibfield  {author} {\bibinfo {author} {\bibfnamefont {T.}~\bibnamefont
  {Yanagida}},\ }\bibfield  {title} {\bibinfo {title} {{Horizontal gauge
  symmetry and masses of neutrinos}},\ }\href@noop {} {\bibfield  {journal}
  {\bibinfo  {journal} {Conf. Proc. C}\ }\textbf {\bibinfo {volume}
  {7902131}},\ \bibinfo {pages} {95} (\bibinfo {year} {1979})}\BibitemShut
  {NoStop}%
\bibitem [{\citenamefont {Schechter}\ and\ \citenamefont
  {Valle}(1980)}]{Schechter:1980gr}%
  \BibitemOpen
  \bibfield  {author} {\bibinfo {author} {\bibfnamefont {J.}~\bibnamefont
  {Schechter}}\ and\ \bibinfo {author} {\bibfnamefont {J.~W.~F.}\ \bibnamefont
  {Valle}},\ }\bibfield  {title} {\bibinfo {title} {{Neutrino Masses in SU(2) x
  U(1) Theories}},\ }\href {https://doi.org/10.1103/PhysRevD.22.2227}
  {\bibfield  {journal} {\bibinfo  {journal} {Phys. Rev. D}\ }\textbf {\bibinfo
  {volume} {22}},\ \bibinfo {pages} {2227} (\bibinfo {year}
  {1980})}\BibitemShut {NoStop}%
\bibitem [{\citenamefont {Schechter}\ and\ \citenamefont
  {Valle}(1982)}]{Schechter:1981cv}%
  \BibitemOpen
  \bibfield  {author} {\bibinfo {author} {\bibfnamefont {J.}~\bibnamefont
  {Schechter}}\ and\ \bibinfo {author} {\bibfnamefont {J.~W.~F.}\ \bibnamefont
  {Valle}},\ }\bibfield  {title} {\bibinfo {title} {{Neutrino Decay and
  Spontaneous Violation of Lepton Number}},\ }\href
  {https://doi.org/10.1103/PhysRevD.25.774} {\bibfield  {journal} {\bibinfo
  {journal} {Phys. Rev. D}\ }\textbf {\bibinfo {volume} {25}},\ \bibinfo
  {pages} {774} (\bibinfo {year} {1982})}\BibitemShut {NoStop}%
\bibitem [{\citenamefont {Datta}\ \emph {et~al.}(2021)\citenamefont {Datta},
  \citenamefont {Roshan},\ and\ \citenamefont {Sil}}]{Datta:2021elq}%
  \BibitemOpen
  \bibfield  {author} {\bibinfo {author} {\bibfnamefont {A.}~\bibnamefont
  {Datta}}, \bibinfo {author} {\bibfnamefont {R.}~\bibnamefont {Roshan}},\ and\
  \bibinfo {author} {\bibfnamefont {A.}~\bibnamefont {Sil}},\ }\bibfield
  {title} {\bibinfo {title} {{Imprint of the Seesaw Mechanism on Feebly
  Interacting Dark Matter and the Baryon Asymmetry}},\ }\href
  {https://doi.org/10.1103/PhysRevLett.127.231801} {\bibfield  {journal}
  {\bibinfo  {journal} {Phys. Rev. Lett.}\ }\textbf {\bibinfo {volume} {127}},\
  \bibinfo {pages} {231801} (\bibinfo {year} {2021})},\ \Eprint
  {https://arxiv.org/abs/2104.02030} {arXiv:2104.02030 [hep-ph]} \BibitemShut
  {NoStop}%
\bibitem [{\citenamefont {{Julian}}(1967)}]{1967ApJ...148..175J}%
  \BibitemOpen
  \bibfield  {author} {\bibinfo {author} {\bibfnamefont {W.~H.}\ \bibnamefont
  {{Julian}}},\ }\bibfield  {title} {\bibinfo {title} {{On the Effect of
  Interstellar Material on Stellar Non-Circular Velocities in Disk Galaxies}},\
  }\href {https://doi.org/10.1086/149134} {\bibfield  {journal} {\bibinfo
  {journal} {\apj}\ }\textbf {\bibinfo {volume} {148}},\ \bibinfo {pages} {175}
  (\bibinfo {year} {1967})}\BibitemShut {NoStop}%
\bibitem [{\citenamefont {Tegmark}\ \emph {et~al.}(2004)\citenamefont {Tegmark}
  \emph {et~al.}}]{SDSS:2003eyi}%
  \BibitemOpen
  \bibfield  {author} {\bibinfo {author} {\bibfnamefont {M.}~\bibnamefont
  {Tegmark}} \emph {et~al.} (\bibinfo {collaboration} {SDSS}),\ }\bibfield
  {title} {\bibinfo {title} {{Cosmological parameters from SDSS and WMAP}},\
  }\href {https://doi.org/10.1103/PhysRevD.69.103501} {\bibfield  {journal}
  {\bibinfo  {journal} {Phys. Rev. D}\ }\textbf {\bibinfo {volume} {69}},\
  \bibinfo {pages} {103501} (\bibinfo {year} {2004})},\ \Eprint
  {https://arxiv.org/abs/astro-ph/0310723} {arXiv:astro-ph/0310723}
  \BibitemShut {NoStop}%
\bibitem [{\citenamefont {Davidson}\ and\ \citenamefont
  {Ibarra}(2002)}]{Davidson:2002qv}%
  \BibitemOpen
  \bibfield  {author} {\bibinfo {author} {\bibfnamefont {S.}~\bibnamefont
  {Davidson}}\ and\ \bibinfo {author} {\bibfnamefont {A.}~\bibnamefont
  {Ibarra}},\ }\bibfield  {title} {\bibinfo {title} {{A Lower bound on the
  right-handed neutrino mass from leptogenesis}},\ }\href
  {https://doi.org/10.1016/S0370-2693(02)01735-5} {\bibfield  {journal}
  {\bibinfo  {journal} {Phys. Lett. B}\ }\textbf {\bibinfo {volume} {535}},\
  \bibinfo {pages} {25} (\bibinfo {year} {2002})},\ \Eprint
  {https://arxiv.org/abs/hep-ph/0202239} {arXiv:hep-ph/0202239} \BibitemShut
  {NoStop}%
\bibitem [{\citenamefont {Flanz}\ \emph {et~al.}(1995)\citenamefont {Flanz},
  \citenamefont {Paschos},\ and\ \citenamefont {Sarkar}}]{Flanz:1994yx}%
  \BibitemOpen
  \bibfield  {author} {\bibinfo {author} {\bibfnamefont {M.}~\bibnamefont
  {Flanz}}, \bibinfo {author} {\bibfnamefont {E.~A.}\ \bibnamefont {Paschos}},\
  and\ \bibinfo {author} {\bibfnamefont {U.}~\bibnamefont {Sarkar}},\
  }\bibfield  {title} {\bibinfo {title} {{Baryogenesis from a lepton asymmetric
  universe}},\ }\href {https://doi.org/10.1016/0370-2693(94)01555-Q} {\bibfield
   {journal} {\bibinfo  {journal} {Phys. Lett. B}\ }\textbf {\bibinfo {volume}
  {345}},\ \bibinfo {pages} {248} (\bibinfo {year} {1995})},\ \bibinfo {note}
  {[Erratum: Phys.Lett.B 384, 487--487 (1996), Erratum: Phys.Lett.B 382,
  447--447 (1996)]},\ \Eprint {https://arxiv.org/abs/hep-ph/9411366}
  {arXiv:hep-ph/9411366} \BibitemShut {NoStop}%
\bibitem [{\citenamefont {Flanz}\ \emph {et~al.}(1996)\citenamefont {Flanz},
  \citenamefont {Paschos}, \citenamefont {Sarkar},\ and\ \citenamefont
  {Weiss}}]{Flanz:1996fb}%
  \BibitemOpen
  \bibfield  {author} {\bibinfo {author} {\bibfnamefont {M.}~\bibnamefont
  {Flanz}}, \bibinfo {author} {\bibfnamefont {E.~A.}\ \bibnamefont {Paschos}},
  \bibinfo {author} {\bibfnamefont {U.}~\bibnamefont {Sarkar}},\ and\ \bibinfo
  {author} {\bibfnamefont {J.}~\bibnamefont {Weiss}},\ }\bibfield  {title}
  {\bibinfo {title} {{Baryogenesis through mixing of heavy Majorana
  neutrinos}},\ }\href {https://doi.org/10.1016/S0370-2693(96)01337-8}
  {\bibfield  {journal} {\bibinfo  {journal} {Phys. Lett. B}\ }\textbf
  {\bibinfo {volume} {389}},\ \bibinfo {pages} {693} (\bibinfo {year}
  {1996})},\ \Eprint {https://arxiv.org/abs/hep-ph/9607310}
  {arXiv:hep-ph/9607310} \BibitemShut {NoStop}%
\bibitem [{\citenamefont {Pilaftsis}(1997)}]{Pilaftsis:1997jf}%
  \BibitemOpen
  \bibfield  {author} {\bibinfo {author} {\bibfnamefont {A.}~\bibnamefont
  {Pilaftsis}},\ }\bibfield  {title} {\bibinfo {title} {{CP violation and
  baryogenesis due to heavy Majorana neutrinos}},\ }\href
  {https://doi.org/10.1103/PhysRevD.56.5431} {\bibfield  {journal} {\bibinfo
  {journal} {Phys. Rev. D}\ }\textbf {\bibinfo {volume} {56}},\ \bibinfo
  {pages} {5431} (\bibinfo {year} {1997})},\ \Eprint
  {https://arxiv.org/abs/hep-ph/9707235} {arXiv:hep-ph/9707235} \BibitemShut
  {NoStop}%
\bibitem [{\citenamefont {Pilaftsis}\ and\ \citenamefont
  {Underwood}(2004)}]{Pilaftsis:2003gt}%
  \BibitemOpen
  \bibfield  {author} {\bibinfo {author} {\bibfnamefont {A.}~\bibnamefont
  {Pilaftsis}}\ and\ \bibinfo {author} {\bibfnamefont {T.~E.~J.}\ \bibnamefont
  {Underwood}},\ }\bibfield  {title} {\bibinfo {title} {{Resonant
  leptogenesis}},\ }\href {https://doi.org/10.1016/j.nuclphysb.2004.05.029}
  {\bibfield  {journal} {\bibinfo  {journal} {Nucl. Phys. B}\ }\textbf
  {\bibinfo {volume} {692}},\ \bibinfo {pages} {303} (\bibinfo {year}
  {2004})},\ \Eprint {https://arxiv.org/abs/hep-ph/0309342}
  {arXiv:hep-ph/0309342} \BibitemShut {NoStop}%
\bibitem [{\citenamefont {D'Onofrio}\ \emph {et~al.}(2014)\citenamefont
  {D'Onofrio}, \citenamefont {Rummukainen},\ and\ \citenamefont
  {Tranberg}}]{DOnofrio:2014rug}%
  \BibitemOpen
  \bibfield  {author} {\bibinfo {author} {\bibfnamefont {M.}~\bibnamefont
  {D'Onofrio}}, \bibinfo {author} {\bibfnamefont {K.}~\bibnamefont
  {Rummukainen}},\ and\ \bibinfo {author} {\bibfnamefont {A.}~\bibnamefont
  {Tranberg}},\ }\bibfield  {title} {\bibinfo {title} {{Sphaleron Rate in the
  Minimal Standard Model}},\ }\href
  {https://doi.org/10.1103/PhysRevLett.113.141602} {\bibfield  {journal}
  {\bibinfo  {journal} {Phys. Rev. Lett.}\ }\textbf {\bibinfo {volume} {113}},\
  \bibinfo {pages} {141602} (\bibinfo {year} {2014})},\ \Eprint
  {https://arxiv.org/abs/1404.3565} {arXiv:1404.3565 [hep-ph]} \BibitemShut
  {NoStop}%
\bibitem [{\citenamefont {D'Onofrio}\ and\ \citenamefont
  {Rummukainen}(2016)}]{DOnofrio:2015gop}%
  \BibitemOpen
  \bibfield  {author} {\bibinfo {author} {\bibfnamefont {M.}~\bibnamefont
  {D'Onofrio}}\ and\ \bibinfo {author} {\bibfnamefont {K.}~\bibnamefont
  {Rummukainen}},\ }\bibfield  {title} {\bibinfo {title} {{Standard model
  cross-over on the lattice}},\ }\href
  {https://doi.org/10.1103/PhysRevD.93.025003} {\bibfield  {journal} {\bibinfo
  {journal} {Phys. Rev. D}\ }\textbf {\bibinfo {volume} {93}},\ \bibinfo
  {pages} {025003} (\bibinfo {year} {2016})},\ \Eprint
  {https://arxiv.org/abs/1508.07161} {arXiv:1508.07161 [hep-ph]} \BibitemShut
  {NoStop}%
\bibitem [{\citenamefont {Hambye}\ and\ \citenamefont
  {Teresi}(2016)}]{Hambye:2016sby}%
  \BibitemOpen
  \bibfield  {author} {\bibinfo {author} {\bibfnamefont {T.}~\bibnamefont
  {Hambye}}\ and\ \bibinfo {author} {\bibfnamefont {D.}~\bibnamefont
  {Teresi}},\ }\bibfield  {title} {\bibinfo {title} {{Higgs doublet decay as
  the origin of the baryon asymmetry}},\ }\href
  {https://doi.org/10.1103/PhysRevLett.117.091801} {\bibfield  {journal}
  {\bibinfo  {journal} {Phys. Rev. Lett.}\ }\textbf {\bibinfo {volume} {117}},\
  \bibinfo {pages} {091801} (\bibinfo {year} {2016})},\ \Eprint
  {https://arxiv.org/abs/1606.00017} {arXiv:1606.00017 [hep-ph]} \BibitemShut
  {NoStop}%
\bibitem [{\citenamefont {Akhmedov}\ \emph {et~al.}(1998)\citenamefont
  {Akhmedov}, \citenamefont {Rubakov},\ and\ \citenamefont
  {Smirnov}}]{Akhmedov:1998qx}%
  \BibitemOpen
  \bibfield  {author} {\bibinfo {author} {\bibfnamefont {E.~K.}\ \bibnamefont
  {Akhmedov}}, \bibinfo {author} {\bibfnamefont {V.~A.}\ \bibnamefont
  {Rubakov}},\ and\ \bibinfo {author} {\bibfnamefont {A.~Y.}\ \bibnamefont
  {Smirnov}},\ }\bibfield  {title} {\bibinfo {title} {{Baryogenesis via
  neutrino oscillations}},\ }\href
  {https://doi.org/10.1103/PhysRevLett.81.1359} {\bibfield  {journal} {\bibinfo
   {journal} {Phys. Rev. Lett.}\ }\textbf {\bibinfo {volume} {81}},\ \bibinfo
  {pages} {1359} (\bibinfo {year} {1998})},\ \Eprint
  {https://arxiv.org/abs/hep-ph/9803255} {arXiv:hep-ph/9803255} \BibitemShut
  {NoStop}%
\bibitem [{\citenamefont {Bhandari}\ \emph {et~al.}(2024)\citenamefont
  {Bhandari}, \citenamefont {Datta},\ and\ \citenamefont
  {Sil}}]{Bhandari:2023wit}%
  \BibitemOpen
  \bibfield  {author} {\bibinfo {author} {\bibfnamefont {D.}~\bibnamefont
  {Bhandari}}, \bibinfo {author} {\bibfnamefont {A.}~\bibnamefont {Datta}},\
  and\ \bibinfo {author} {\bibfnamefont {A.}~\bibnamefont {Sil}},\ }\bibfield
  {title} {\bibinfo {title} {{Leptogenesis from a phase transition in a
  dynamical vacuum}},\ }\href {https://doi.org/10.1103/PhysRevD.110.115008}
  {\bibfield  {journal} {\bibinfo  {journal} {Phys. Rev. D}\ }\textbf {\bibinfo
  {volume} {110}},\ \bibinfo {pages} {115008} (\bibinfo {year} {2024})},\
  \Eprint {https://arxiv.org/abs/2312.13157} {arXiv:2312.13157 [hep-ph]}
  \BibitemShut {NoStop}%
\bibitem [{\citenamefont {Cline}(2006)}]{Cline:2006ts}%
  \BibitemOpen
  \bibfield  {author} {\bibinfo {author} {\bibfnamefont {J.~M.}\ \bibnamefont
  {Cline}},\ }\bibfield  {title} {\bibinfo {title} {{Baryogenesis}},\ }in\
  \href@noop {} {\emph {\bibinfo {booktitle} {{Les Houches Summer School -
  Session 86: Particle Physics and Cosmology: The Fabric of Spacetime}}}}\
  (\bibinfo {year} {2006})\ \Eprint {https://arxiv.org/abs/hep-ph/0609145}
  {arXiv:hep-ph/0609145} \BibitemShut {NoStop}%
\bibitem [{\citenamefont {Guth}\ and\ \citenamefont
  {Weinberg}(1981)}]{Guth:1981uk}%
  \BibitemOpen
  \bibfield  {author} {\bibinfo {author} {\bibfnamefont {A.~H.}\ \bibnamefont
  {Guth}}\ and\ \bibinfo {author} {\bibfnamefont {E.~J.}\ \bibnamefont
  {Weinberg}},\ }\bibfield  {title} {\bibinfo {title} {{Cosmological
  Consequences of a First Order Phase Transition in the SU(5) Grand Unified
  Model}},\ }\href {https://doi.org/10.1103/PhysRevD.23.876} {\bibfield
  {journal} {\bibinfo  {journal} {Phys. Rev. D}\ }\textbf {\bibinfo {volume}
  {23}},\ \bibinfo {pages} {876} (\bibinfo {year} {1981})}\BibitemShut
  {NoStop}%
\bibitem [{\citenamefont {Guth}\ and\ \citenamefont {Tye}(1980)}]{Guth:1979bh}%
  \BibitemOpen
  \bibfield  {author} {\bibinfo {author} {\bibfnamefont {A.~H.}\ \bibnamefont
  {Guth}}\ and\ \bibinfo {author} {\bibfnamefont {S.~H.~H.}\ \bibnamefont
  {Tye}},\ }\bibfield  {title} {\bibinfo {title} {{Phase Transitions and
  Magnetic Monopole Production in the Very Early Universe}},\ }\href
  {https://doi.org/10.1103/PhysRevLett.44.631} {\bibfield  {journal} {\bibinfo
  {journal} {Phys. Rev. Lett.}\ }\textbf {\bibinfo {volume} {44}},\ \bibinfo
  {pages} {631} (\bibinfo {year} {1980})},\ \bibinfo {note} {[Erratum:
  Phys.Rev.Lett. 44, 963 (1980)]}\BibitemShut {NoStop}%
\bibitem [{\citenamefont {Coleman}(1977)}]{Coleman:1977py}%
  \BibitemOpen
  \bibfield  {author} {\bibinfo {author} {\bibfnamefont {S.~R.}\ \bibnamefont
  {Coleman}},\ }\bibfield  {title} {\bibinfo {title} {{The Fate of the False
  Vacuum. 1. Semiclassical Theory}},\ }\href
  {https://doi.org/10.1103/PhysRevD.16.1248} {\bibfield  {journal} {\bibinfo
  {journal} {Phys. Rev. D}\ }\textbf {\bibinfo {volume} {15}},\ \bibinfo
  {pages} {2929} (\bibinfo {year} {1977})},\ \bibinfo {note} {[Erratum:
  Phys.Rev.D 16, 1248 (1977)]}\BibitemShut {NoStop}%
\bibitem [{\citenamefont {Linde}(1981)}]{Linde:1980tt}%
  \BibitemOpen
  \bibfield  {author} {\bibinfo {author} {\bibfnamefont {A.~D.}\ \bibnamefont
  {Linde}},\ }\bibfield  {title} {\bibinfo {title} {{Fate of the False Vacuum
  at Finite Temperature: Theory and Applications}},\ }\href
  {https://doi.org/10.1016/0370-2693(81)90281-1} {\bibfield  {journal}
  {\bibinfo  {journal} {Phys. Lett. B}\ }\textbf {\bibinfo {volume} {100}},\
  \bibinfo {pages} {37} (\bibinfo {year} {1981})}\BibitemShut {NoStop}%
\bibitem [{\citenamefont {Linde}(1983)}]{Linde:1981zj}%
  \BibitemOpen
  \bibfield  {author} {\bibinfo {author} {\bibfnamefont {A.~D.}\ \bibnamefont
  {Linde}},\ }\bibfield  {title} {\bibinfo {title} {{Decay of the False Vacuum
  at Finite Temperature}},\ }\href
  {https://doi.org/10.1016/0550-3213(83)90072-X} {\bibfield  {journal}
  {\bibinfo  {journal} {Nucl. Phys. B}\ }\textbf {\bibinfo {volume} {216}},\
  \bibinfo {pages} {421} (\bibinfo {year} {1983})},\ \bibinfo {note} {[Erratum:
  Nucl.Phys.B 223, 544 (1983)]}\BibitemShut {NoStop}%
\bibitem [{\citenamefont {Guada}\ \emph {et~al.}(2020)\citenamefont {Guada},
  \citenamefont {Nemev\v{s}ek},\ and\ \citenamefont {Pintar}}]{Guada:2020xnz}%
  \BibitemOpen
  \bibfield  {author} {\bibinfo {author} {\bibfnamefont {V.}~\bibnamefont
  {Guada}}, \bibinfo {author} {\bibfnamefont {M.}~\bibnamefont
  {Nemev\v{s}ek}},\ and\ \bibinfo {author} {\bibfnamefont {M.}~\bibnamefont
  {Pintar}},\ }\bibfield  {title} {\bibinfo {title} {{FindBounce: Package for
  multi-field bounce actions}},\ }\href
  {https://doi.org/10.1016/j.cpc.2020.107480} {\bibfield  {journal} {\bibinfo
  {journal} {Comput. Phys. Commun.}\ }\textbf {\bibinfo {volume} {256}},\
  \bibinfo {pages} {107480} (\bibinfo {year} {2020})},\ \Eprint
  {https://arxiv.org/abs/2002.00881} {arXiv:2002.00881 [hep-ph]} \BibitemShut
  {NoStop}%
\bibitem [{\citenamefont {Espinosa}\ \emph {et~al.}(2012)\citenamefont
  {Espinosa}, \citenamefont {Konstandin},\ and\ \citenamefont
  {Riva}}]{Espinosa:2011ax}%
  \BibitemOpen
  \bibfield  {author} {\bibinfo {author} {\bibfnamefont {J.~R.}\ \bibnamefont
  {Espinosa}}, \bibinfo {author} {\bibfnamefont {T.}~\bibnamefont
  {Konstandin}},\ and\ \bibinfo {author} {\bibfnamefont {F.}~\bibnamefont
  {Riva}},\ }\bibfield  {title} {\bibinfo {title} {{Strong Electroweak Phase
  Transitions in the Standard Model with a Singlet}},\ }\href
  {https://doi.org/10.1016/j.nuclphysb.2011.09.010} {\bibfield  {journal}
  {\bibinfo  {journal} {Nucl. Phys. B}\ }\textbf {\bibinfo {volume} {854}},\
  \bibinfo {pages} {592} (\bibinfo {year} {2012})},\ \Eprint
  {https://arxiv.org/abs/1107.5441} {arXiv:1107.5441 [hep-ph]} \BibitemShut
  {NoStop}%
\bibitem [{\citenamefont {Profumo}\ \emph {et~al.}(2007)\citenamefont
  {Profumo}, \citenamefont {Ramsey-Musolf},\ and\ \citenamefont
  {Shaughnessy}}]{Profumo:2007wc}%
  \BibitemOpen
  \bibfield  {author} {\bibinfo {author} {\bibfnamefont {S.}~\bibnamefont
  {Profumo}}, \bibinfo {author} {\bibfnamefont {M.~J.}\ \bibnamefont
  {Ramsey-Musolf}},\ and\ \bibinfo {author} {\bibfnamefont {G.}~\bibnamefont
  {Shaughnessy}},\ }\bibfield  {title} {\bibinfo {title} {{Singlet Higgs
  phenomenology and the electroweak phase transition}},\ }\href
  {https://doi.org/10.1088/1126-6708/2007/08/010} {\bibfield  {journal}
  {\bibinfo  {journal} {JHEP}\ }\textbf {\bibinfo {volume} {08}},\ \bibinfo
  {pages} {010}},\ \Eprint {https://arxiv.org/abs/0705.2425} {arXiv:0705.2425
  [hep-ph]} \BibitemShut {NoStop}%
\bibitem [{\citenamefont {Davoudiasl}\ \emph {et~al.}(2013)\citenamefont
  {Davoudiasl}, \citenamefont {Lewis},\ and\ \citenamefont
  {Ponton}}]{Davoudiasl:2012tu}%
  \BibitemOpen
  \bibfield  {author} {\bibinfo {author} {\bibfnamefont {H.}~\bibnamefont
  {Davoudiasl}}, \bibinfo {author} {\bibfnamefont {I.}~\bibnamefont {Lewis}},\
  and\ \bibinfo {author} {\bibfnamefont {E.}~\bibnamefont {Ponton}},\
  }\bibfield  {title} {\bibinfo {title} {{Electroweak Phase Transition, Higgs
  Diphoton Rate, and New Heavy Fermions}},\ }\href
  {https://doi.org/10.1103/PhysRevD.87.093001} {\bibfield  {journal} {\bibinfo
  {journal} {Phys. Rev. D}\ }\textbf {\bibinfo {volume} {87}},\ \bibinfo
  {pages} {093001} (\bibinfo {year} {2013})},\ \Eprint
  {https://arxiv.org/abs/1211.3449} {arXiv:1211.3449 [hep-ph]} \BibitemShut
  {NoStop}%
\bibitem [{\citenamefont {Chala}\ \emph {et~al.}(2018)\citenamefont {Chala},
  \citenamefont {Krause},\ and\ \citenamefont {Nardini}}]{Chala:2018ari}%
  \BibitemOpen
  \bibfield  {author} {\bibinfo {author} {\bibfnamefont {M.}~\bibnamefont
  {Chala}}, \bibinfo {author} {\bibfnamefont {C.}~\bibnamefont {Krause}},\ and\
  \bibinfo {author} {\bibfnamefont {G.}~\bibnamefont {Nardini}},\ }\bibfield
  {title} {\bibinfo {title} {{Signals of the electroweak phase transition at
  colliders and gravitational wave observatories}},\ }\href
  {https://doi.org/10.1007/JHEP07(2018)062} {\bibfield  {journal} {\bibinfo
  {journal} {JHEP}\ }\textbf {\bibinfo {volume} {07}},\ \bibinfo {pages}
  {062}},\ \Eprint {https://arxiv.org/abs/1802.02168} {arXiv:1802.02168
  [hep-ph]} \BibitemShut {NoStop}%
\bibitem [{\citenamefont {Bodeker}\ \emph {et~al.}(2005)\citenamefont
  {Bodeker}, \citenamefont {Fromme}, \citenamefont {Huber},\ and\ \citenamefont
  {Seniuch}}]{Bodeker:2004ws}%
  \BibitemOpen
  \bibfield  {author} {\bibinfo {author} {\bibfnamefont {D.}~\bibnamefont
  {Bodeker}}, \bibinfo {author} {\bibfnamefont {L.}~\bibnamefont {Fromme}},
  \bibinfo {author} {\bibfnamefont {S.~J.}\ \bibnamefont {Huber}},\ and\
  \bibinfo {author} {\bibfnamefont {M.}~\bibnamefont {Seniuch}},\ }\bibfield
  {title} {\bibinfo {title} {{The Baryon asymmetry in the standard model with a
  low cut-off}},\ }\href {https://doi.org/10.1088/1126-6708/2005/02/026}
  {\bibfield  {journal} {\bibinfo  {journal} {JHEP}\ }\textbf {\bibinfo
  {volume} {02}},\ \bibinfo {pages} {026}},\ \Eprint
  {https://arxiv.org/abs/hep-ph/0412366} {arXiv:hep-ph/0412366} \BibitemShut
  {NoStop}%
\bibitem [{\citenamefont {Delaunay}\ \emph {et~al.}(2008)\citenamefont
  {Delaunay}, \citenamefont {Grojean},\ and\ \citenamefont
  {Wells}}]{Delaunay:2007wb}%
  \BibitemOpen
  \bibfield  {author} {\bibinfo {author} {\bibfnamefont {C.}~\bibnamefont
  {Delaunay}}, \bibinfo {author} {\bibfnamefont {C.}~\bibnamefont {Grojean}},\
  and\ \bibinfo {author} {\bibfnamefont {J.~D.}\ \bibnamefont {Wells}},\
  }\bibfield  {title} {\bibinfo {title} {{Dynamics of Non-renormalizable
  Electroweak Symmetry Breaking}},\ }\href
  {https://doi.org/10.1088/1126-6708/2008/04/029} {\bibfield  {journal}
  {\bibinfo  {journal} {JHEP}\ }\textbf {\bibinfo {volume} {04}},\ \bibinfo
  {pages} {029}},\ \Eprint {https://arxiv.org/abs/0711.2511} {arXiv:0711.2511
  [hep-ph]} \BibitemShut {NoStop}%
\bibitem [{\citenamefont {Grojean}\ \emph {et~al.}(2005)\citenamefont
  {Grojean}, \citenamefont {Servant},\ and\ \citenamefont
  {Wells}}]{Grojean:2004xa}%
  \BibitemOpen
  \bibfield  {author} {\bibinfo {author} {\bibfnamefont {C.}~\bibnamefont
  {Grojean}}, \bibinfo {author} {\bibfnamefont {G.}~\bibnamefont {Servant}},\
  and\ \bibinfo {author} {\bibfnamefont {J.~D.}\ \bibnamefont {Wells}},\
  }\bibfield  {title} {\bibinfo {title} {{First-order electroweak phase
  transition in the standard model with a low cutoff}},\ }\href
  {https://doi.org/10.1103/PhysRevD.71.036001} {\bibfield  {journal} {\bibinfo
  {journal} {Phys. Rev. D}\ }\textbf {\bibinfo {volume} {71}},\ \bibinfo
  {pages} {036001} (\bibinfo {year} {2005})},\ \Eprint
  {https://arxiv.org/abs/hep-ph/0407019} {arXiv:hep-ph/0407019} \BibitemShut
  {NoStop}%
\bibitem [{\citenamefont {Huang}\ \emph
  {et~al.}(2016{\natexlab{a}})\citenamefont {Huang}, \citenamefont {Gu},
  \citenamefont {Yin}, \citenamefont {Yu},\ and\ \citenamefont
  {Zhang}}]{Huang:2015izx}%
  \BibitemOpen
  \bibfield  {author} {\bibinfo {author} {\bibfnamefont {F.~P.}\ \bibnamefont
  {Huang}}, \bibinfo {author} {\bibfnamefont {P.-H.}\ \bibnamefont {Gu}},
  \bibinfo {author} {\bibfnamefont {P.-F.}\ \bibnamefont {Yin}}, \bibinfo
  {author} {\bibfnamefont {Z.-H.}\ \bibnamefont {Yu}},\ and\ \bibinfo {author}
  {\bibfnamefont {X.}~\bibnamefont {Zhang}},\ }\bibfield  {title} {\bibinfo
  {title} {{Testing the electroweak phase transition and electroweak
  baryogenesis at the LHC and a circular electron-positron collider}},\ }\href
  {https://doi.org/10.1103/PhysRevD.93.103515} {\bibfield  {journal} {\bibinfo
  {journal} {Phys. Rev. D}\ }\textbf {\bibinfo {volume} {93}},\ \bibinfo
  {pages} {103515} (\bibinfo {year} {2016}{\natexlab{a}})},\ \Eprint
  {https://arxiv.org/abs/1511.03969} {arXiv:1511.03969 [hep-ph]} \BibitemShut
  {NoStop}%
\bibitem [{\citenamefont {Huang}\ \emph
  {et~al.}(2016{\natexlab{b}})\citenamefont {Huang}, \citenamefont {Wan},
  \citenamefont {Wang}, \citenamefont {Cai},\ and\ \citenamefont
  {Zhang}}]{Huang:2016odd}%
  \BibitemOpen
  \bibfield  {author} {\bibinfo {author} {\bibfnamefont {F.~P.}\ \bibnamefont
  {Huang}}, \bibinfo {author} {\bibfnamefont {Y.}~\bibnamefont {Wan}}, \bibinfo
  {author} {\bibfnamefont {D.-G.}\ \bibnamefont {Wang}}, \bibinfo {author}
  {\bibfnamefont {Y.-F.}\ \bibnamefont {Cai}},\ and\ \bibinfo {author}
  {\bibfnamefont {X.}~\bibnamefont {Zhang}},\ }\bibfield  {title} {\bibinfo
  {title} {{Hearing the echoes of electroweak baryogenesis with gravitational
  wave detectors}},\ }\href {https://doi.org/10.1103/PhysRevD.94.041702}
  {\bibfield  {journal} {\bibinfo  {journal} {Phys. Rev. D}\ }\textbf {\bibinfo
  {volume} {94}},\ \bibinfo {pages} {041702} (\bibinfo {year}
  {2016}{\natexlab{b}})},\ \Eprint {https://arxiv.org/abs/1601.01640}
  {arXiv:1601.01640 [hep-ph]} \BibitemShut {NoStop}%
\bibitem [{\citenamefont {Espinosa}\ \emph {et~al.}(1992)\citenamefont
  {Espinosa}, \citenamefont {Quiros},\ and\ \citenamefont
  {Zwirner}}]{Espinosa:1992gq}%
  \BibitemOpen
  \bibfield  {author} {\bibinfo {author} {\bibfnamefont {J.~R.}\ \bibnamefont
  {Espinosa}}, \bibinfo {author} {\bibfnamefont {M.}~\bibnamefont {Quiros}},\
  and\ \bibinfo {author} {\bibfnamefont {F.}~\bibnamefont {Zwirner}},\
  }\bibfield  {title} {\bibinfo {title} {{On the phase transition in the scalar
  theory}},\ }\href {https://doi.org/10.1016/0370-2693(92)90129-R} {\bibfield
  {journal} {\bibinfo  {journal} {Phys. Lett. B}\ }\textbf {\bibinfo {volume}
  {291}},\ \bibinfo {pages} {115} (\bibinfo {year} {1992})},\ \Eprint
  {https://arxiv.org/abs/hep-ph/9206227} {arXiv:hep-ph/9206227} \BibitemShut
  {NoStop}%
\bibitem [{\citenamefont {Wainwright}(2012)}]{Wainwright:2011kj}%
  \BibitemOpen
  \bibfield  {author} {\bibinfo {author} {\bibfnamefont {C.~L.}\ \bibnamefont
  {Wainwright}},\ }\bibfield  {title} {\bibinfo {title} {{CosmoTransitions:
  Computing Cosmological Phase Transition Temperatures and Bubble Profiles with
  Multiple Fields}},\ }\href {https://doi.org/10.1016/j.cpc.2012.04.004}
  {\bibfield  {journal} {\bibinfo  {journal} {Comput. Phys. Commun.}\ }\textbf
  {\bibinfo {volume} {183}},\ \bibinfo {pages} {2006} (\bibinfo {year}
  {2012})},\ \Eprint {https://arxiv.org/abs/1109.4189} {arXiv:1109.4189
  [hep-ph]} \BibitemShut {NoStop}%
\bibitem [{\citenamefont {Ai}\ \emph {et~al.}(2023)\citenamefont {Ai},
  \citenamefont {Laurent},\ and\ \citenamefont {van~de Vis}}]{Ai:2023see}%
  \BibitemOpen
  \bibfield  {author} {\bibinfo {author} {\bibfnamefont {W.-Y.}\ \bibnamefont
  {Ai}}, \bibinfo {author} {\bibfnamefont {B.}~\bibnamefont {Laurent}},\ and\
  \bibinfo {author} {\bibfnamefont {J.}~\bibnamefont {van~de Vis}},\ }\bibfield
   {title} {\bibinfo {title} {{Model-independent bubble wall velocities in
  local thermal equilibrium}},\ }\href
  {https://doi.org/10.1088/1475-7516/2023/07/002} {\bibfield  {journal}
  {\bibinfo  {journal} {JCAP}\ }\textbf {\bibinfo {volume} {07}},\ \bibinfo
  {pages} {002}},\ \Eprint {https://arxiv.org/abs/2303.10171} {arXiv:2303.10171
  [astro-ph.CO]} \BibitemShut {NoStop}%
\bibitem [{\citenamefont {Ai}\ \emph {et~al.}(2025)\citenamefont {Ai},
  \citenamefont {Laurent},\ and\ \citenamefont {van~de Vis}}]{Ai:2024btx}%
  \BibitemOpen
  \bibfield  {author} {\bibinfo {author} {\bibfnamefont {W.-Y.}\ \bibnamefont
  {Ai}}, \bibinfo {author} {\bibfnamefont {B.}~\bibnamefont {Laurent}},\ and\
  \bibinfo {author} {\bibfnamefont {J.}~\bibnamefont {van~de Vis}},\ }\bibfield
   {title} {\bibinfo {title} {{Bounds on the bubble wall velocity}},\ }\href
  {https://doi.org/10.1007/JHEP02(2025)119} {\bibfield  {journal} {\bibinfo
  {journal} {JHEP}\ }\textbf {\bibinfo {volume} {02}},\ \bibinfo {pages}
  {119}},\ \Eprint {https://arxiv.org/abs/2411.13641} {arXiv:2411.13641
  [hep-ph]} \BibitemShut {NoStop}%
\bibitem [{\citenamefont {Carena}\ \emph {et~al.}(2023)\citenamefont {Carena},
  \citenamefont {Li}, \citenamefont {Ou},\ and\ \citenamefont
  {Wang}}]{Carena:2022qpf}%
  \BibitemOpen
  \bibfield  {author} {\bibinfo {author} {\bibfnamefont {M.}~\bibnamefont
  {Carena}}, \bibinfo {author} {\bibfnamefont {Y.-Y.}\ \bibnamefont {Li}},
  \bibinfo {author} {\bibfnamefont {T.}~\bibnamefont {Ou}},\ and\ \bibinfo
  {author} {\bibfnamefont {Y.}~\bibnamefont {Wang}},\ }\bibfield  {title}
  {\bibinfo {title} {{Anatomy of the electroweak phase transition for dark
  sector induced baryogenesis}},\ }\href
  {https://doi.org/10.1007/JHEP02(2023)139} {\bibfield  {journal} {\bibinfo
  {journal} {JHEP}\ }\textbf {\bibinfo {volume} {02}},\ \bibinfo {pages}
  {139}},\ \Eprint {https://arxiv.org/abs/2210.14352} {arXiv:2210.14352
  [hep-ph]} \BibitemShut {NoStop}%
\bibitem [{\citenamefont {Datta}\ \emph {et~al.}(2024)\citenamefont {Datta},
  \citenamefont {Roshan},\ and\ \citenamefont {Sil}}]{Datta:2022jic}%
  \BibitemOpen
  \bibfield  {author} {\bibinfo {author} {\bibfnamefont {A.}~\bibnamefont
  {Datta}}, \bibinfo {author} {\bibfnamefont {R.}~\bibnamefont {Roshan}},\ and\
  \bibinfo {author} {\bibfnamefont {A.}~\bibnamefont {Sil}},\ }\bibfield
  {title} {\bibinfo {title} {{Effects of Reheating on Charged Lepton Yukawa
  Equilibration and Leptogenesis}},\ }\href
  {https://doi.org/10.1103/PhysRevLett.132.061802} {\bibfield  {journal}
  {\bibinfo  {journal} {Phys. Rev. Lett.}\ }\textbf {\bibinfo {volume} {132}},\
  \bibinfo {pages} {061802} (\bibinfo {year} {2024})},\ \Eprint
  {https://arxiv.org/abs/2206.10650} {arXiv:2206.10650 [hep-ph]} \BibitemShut
  {NoStop}%
\bibitem [{\citenamefont {Datta}\ \emph {et~al.}(2023)\citenamefont {Datta},
  \citenamefont {Roshan},\ and\ \citenamefont {Sil}}]{Datta:2023pav}%
  \BibitemOpen
  \bibfield  {author} {\bibinfo {author} {\bibfnamefont {A.}~\bibnamefont
  {Datta}}, \bibinfo {author} {\bibfnamefont {R.}~\bibnamefont {Roshan}},\ and\
  \bibinfo {author} {\bibfnamefont {A.}~\bibnamefont {Sil}},\ }\bibfield
  {title} {\bibinfo {title} {{Flavor leptogenesis during the reheating era}},\
  }\href {https://doi.org/10.1103/PhysRevD.108.035029} {\bibfield  {journal}
  {\bibinfo  {journal} {Phys. Rev. D}\ }\textbf {\bibinfo {volume} {108}},\
  \bibinfo {pages} {035029} (\bibinfo {year} {2023})},\ \Eprint
  {https://arxiv.org/abs/2301.10791} {arXiv:2301.10791 [hep-ph]} \BibitemShut
  {NoStop}%
\bibitem [{\citenamefont {Kawasaki}\ \emph {et~al.}(1999)\citenamefont
  {Kawasaki}, \citenamefont {Kohri},\ and\ \citenamefont
  {Sugiyama}}]{Kawasaki:1999na}%
  \BibitemOpen
  \bibfield  {author} {\bibinfo {author} {\bibfnamefont {M.}~\bibnamefont
  {Kawasaki}}, \bibinfo {author} {\bibfnamefont {K.}~\bibnamefont {Kohri}},\
  and\ \bibinfo {author} {\bibfnamefont {N.}~\bibnamefont {Sugiyama}},\
  }\bibfield  {title} {\bibinfo {title} {{Cosmological constraints on late time
  entropy production}},\ }\href {https://doi.org/10.1103/PhysRevLett.82.4168}
  {\bibfield  {journal} {\bibinfo  {journal} {Phys. Rev. Lett.}\ }\textbf
  {\bibinfo {volume} {82}},\ \bibinfo {pages} {4168} (\bibinfo {year}
  {1999})},\ \Eprint {https://arxiv.org/abs/astro-ph/9811437}
  {arXiv:astro-ph/9811437} \BibitemShut {NoStop}%
\bibitem [{\citenamefont {Kawasaki}\ \emph {et~al.}(2000)\citenamefont
  {Kawasaki}, \citenamefont {Kohri},\ and\ \citenamefont
  {Sugiyama}}]{Kawasaki:2000en}%
  \BibitemOpen
  \bibfield  {author} {\bibinfo {author} {\bibfnamefont {M.}~\bibnamefont
  {Kawasaki}}, \bibinfo {author} {\bibfnamefont {K.}~\bibnamefont {Kohri}},\
  and\ \bibinfo {author} {\bibfnamefont {N.}~\bibnamefont {Sugiyama}},\
  }\bibfield  {title} {\bibinfo {title} {{MeV scale reheating temperature and
  thermalization of neutrino background}},\ }\href
  {https://doi.org/10.1103/PhysRevD.62.023506} {\bibfield  {journal} {\bibinfo
  {journal} {Phys. Rev. D}\ }\textbf {\bibinfo {volume} {62}},\ \bibinfo
  {pages} {023506} (\bibinfo {year} {2000})},\ \Eprint
  {https://arxiv.org/abs/astro-ph/0002127} {arXiv:astro-ph/0002127}
  \BibitemShut {NoStop}%
\bibitem [{\citenamefont {Giudice}\ \emph
  {et~al.}(2001{\natexlab{a}})\citenamefont {Giudice}, \citenamefont {Kolb},
  \citenamefont {Riotto}, \citenamefont {Semikoz},\ and\ \citenamefont
  {Tkachev}}]{Giudice:2000dp}%
  \BibitemOpen
  \bibfield  {author} {\bibinfo {author} {\bibfnamefont {G.~F.}\ \bibnamefont
  {Giudice}}, \bibinfo {author} {\bibfnamefont {E.~W.}\ \bibnamefont {Kolb}},
  \bibinfo {author} {\bibfnamefont {A.}~\bibnamefont {Riotto}}, \bibinfo
  {author} {\bibfnamefont {D.~V.}\ \bibnamefont {Semikoz}},\ and\ \bibinfo
  {author} {\bibfnamefont {I.~I.}\ \bibnamefont {Tkachev}},\ }\bibfield
  {title} {\bibinfo {title} {{Standard model neutrinos as warm dark matter}},\
  }\href {https://doi.org/10.1103/PhysRevD.64.043512} {\bibfield  {journal}
  {\bibinfo  {journal} {Phys. Rev. D}\ }\textbf {\bibinfo {volume} {64}},\
  \bibinfo {pages} {043512} (\bibinfo {year} {2001}{\natexlab{a}})},\ \Eprint
  {https://arxiv.org/abs/hep-ph/0012317} {arXiv:hep-ph/0012317} \BibitemShut
  {NoStop}%
\bibitem [{\citenamefont {Giudice}\ \emph
  {et~al.}(2001{\natexlab{b}})\citenamefont {Giudice}, \citenamefont {Kolb},\
  and\ \citenamefont {Riotto}}]{Giudice:2000ex}%
  \BibitemOpen
  \bibfield  {author} {\bibinfo {author} {\bibfnamefont {G.~F.}\ \bibnamefont
  {Giudice}}, \bibinfo {author} {\bibfnamefont {E.~W.}\ \bibnamefont {Kolb}},\
  and\ \bibinfo {author} {\bibfnamefont {A.}~\bibnamefont {Riotto}},\
  }\bibfield  {title} {\bibinfo {title} {{Largest temperature of the radiation
  era and its cosmological implications}},\ }\href
  {https://doi.org/10.1103/PhysRevD.64.023508} {\bibfield  {journal} {\bibinfo
  {journal} {Phys. Rev. D}\ }\textbf {\bibinfo {volume} {64}},\ \bibinfo
  {pages} {023508} (\bibinfo {year} {2001}{\natexlab{b}})},\ \Eprint
  {https://arxiv.org/abs/hep-ph/0005123} {arXiv:hep-ph/0005123} \BibitemShut
  {NoStop}%
\bibitem [{\citenamefont {Casas}\ and\ \citenamefont
  {Ibarra}(2001)}]{Casas:2001sr}%
  \BibitemOpen
  \bibfield  {author} {\bibinfo {author} {\bibfnamefont {J.~A.}\ \bibnamefont
  {Casas}}\ and\ \bibinfo {author} {\bibfnamefont {A.}~\bibnamefont {Ibarra}},\
  }\bibfield  {title} {\bibinfo {title} {{Oscillating neutrinos and $\mu \to e,
  \gamma$}},\ }\href {https://doi.org/10.1016/S0550-3213(01)00475-8} {\bibfield
   {journal} {\bibinfo  {journal} {Nucl. Phys. B}\ }\textbf {\bibinfo {volume}
  {618}},\ \bibinfo {pages} {171} (\bibinfo {year} {2001})},\ \Eprint
  {https://arxiv.org/abs/hep-ph/0103065} {arXiv:hep-ph/0103065} \BibitemShut
  {NoStop}%
\bibitem [{\citenamefont {Zyla}\ \emph {et~al.}(2020)\citenamefont {Zyla} \emph
  {et~al.}}]{Zyla:2020zbs}%
  \BibitemOpen
  \bibfield  {author} {\bibinfo {author} {\bibfnamefont {P.~A.}\ \bibnamefont
  {Zyla}} \emph {et~al.} (\bibinfo {collaboration} {Particle Data Group}),\
  }\bibfield  {title} {\bibinfo {title} {{Review of Particle Physics}},\ }\href
  {https://doi.org/10.1093/ptep/ptaa104} {\bibfield  {journal} {\bibinfo
  {journal} {PTEP}\ }\textbf {\bibinfo {volume} {2020}},\ \bibinfo {pages}
  {083C01} (\bibinfo {year} {2020})}\BibitemShut {NoStop}%
\bibitem [{\citenamefont {Davidson}\ \emph {et~al.}(2008)\citenamefont
  {Davidson}, \citenamefont {Nardi},\ and\ \citenamefont
  {Nir}}]{Davidson:2008bu}%
  \BibitemOpen
  \bibfield  {author} {\bibinfo {author} {\bibfnamefont {S.}~\bibnamefont
  {Davidson}}, \bibinfo {author} {\bibfnamefont {E.}~\bibnamefont {Nardi}},\
  and\ \bibinfo {author} {\bibfnamefont {Y.}~\bibnamefont {Nir}},\ }\bibfield
  {title} {\bibinfo {title} {{Leptogenesis}},\ }\href
  {https://doi.org/10.1016/j.physrep.2008.06.002} {\bibfield  {journal}
  {\bibinfo  {journal} {Phys. Rept.}\ }\textbf {\bibinfo {volume} {466}},\
  \bibinfo {pages} {105} (\bibinfo {year} {2008})},\ \Eprint
  {https://arxiv.org/abs/0802.2962} {arXiv:0802.2962 [hep-ph]} \BibitemShut
  {NoStop}%
\bibitem [{\citenamefont {Giudice}\ \emph {et~al.}(2004)\citenamefont
  {Giudice}, \citenamefont {Notari}, \citenamefont {Raidal}, \citenamefont
  {Riotto},\ and\ \citenamefont {Strumia}}]{Giudice:2003jh}%
  \BibitemOpen
  \bibfield  {author} {\bibinfo {author} {\bibfnamefont {G.~F.}\ \bibnamefont
  {Giudice}}, \bibinfo {author} {\bibfnamefont {A.}~\bibnamefont {Notari}},
  \bibinfo {author} {\bibfnamefont {M.}~\bibnamefont {Raidal}}, \bibinfo
  {author} {\bibfnamefont {A.}~\bibnamefont {Riotto}},\ and\ \bibinfo {author}
  {\bibfnamefont {A.}~\bibnamefont {Strumia}},\ }\bibfield  {title} {\bibinfo
  {title} {{Towards a complete theory of thermal leptogenesis in the SM and
  MSSM}},\ }\href {https://doi.org/10.1016/j.nuclphysb.2004.02.019} {\bibfield
  {journal} {\bibinfo  {journal} {Nucl. Phys. B}\ }\textbf {\bibinfo {volume}
  {685}},\ \bibinfo {pages} {89} (\bibinfo {year} {2004})},\ \Eprint
  {https://arxiv.org/abs/hep-ph/0310123} {arXiv:hep-ph/0310123} \BibitemShut
  {NoStop}%
\bibitem [{\citenamefont {Pramanick}\ \emph {et~al.}(2024)\citenamefont
  {Pramanick}, \citenamefont {Ray},\ and\ \citenamefont
  {Sil}}]{Pramanick:2024gvu}%
  \BibitemOpen
  \bibfield  {author} {\bibinfo {author} {\bibfnamefont {R.}~\bibnamefont
  {Pramanick}}, \bibinfo {author} {\bibfnamefont {T.~S.}\ \bibnamefont {Ray}},\
  and\ \bibinfo {author} {\bibfnamefont {A.}~\bibnamefont {Sil}},\ }\bibfield
  {title} {\bibinfo {title} {{Toward a more complete description of hybrid
  leptogenesis}},\ }\href {https://doi.org/10.1103/PhysRevD.109.115011}
  {\bibfield  {journal} {\bibinfo  {journal} {Phys. Rev. D}\ }\textbf {\bibinfo
  {volume} {109}},\ \bibinfo {pages} {115011} (\bibinfo {year} {2024})},\
  \Eprint {https://arxiv.org/abs/2401.12189} {arXiv:2401.12189 [hep-ph]}
  \BibitemShut {NoStop}%
\bibitem [{\citenamefont {Antusch}\ \emph {et~al.}(2018)\citenamefont
  {Antusch}, \citenamefont {Cazzato}, \citenamefont {Drewes}, \citenamefont
  {Fischer}, \citenamefont {Garbrecht}, \citenamefont {Gueter},\ and\
  \citenamefont {Klaric}}]{Antusch:2017pkq}%
  \BibitemOpen
  \bibfield  {author} {\bibinfo {author} {\bibfnamefont {S.}~\bibnamefont
  {Antusch}}, \bibinfo {author} {\bibfnamefont {E.}~\bibnamefont {Cazzato}},
  \bibinfo {author} {\bibfnamefont {M.}~\bibnamefont {Drewes}}, \bibinfo
  {author} {\bibfnamefont {O.}~\bibnamefont {Fischer}}, \bibinfo {author}
  {\bibfnamefont {B.}~\bibnamefont {Garbrecht}}, \bibinfo {author}
  {\bibfnamefont {D.}~\bibnamefont {Gueter}},\ and\ \bibinfo {author}
  {\bibfnamefont {J.}~\bibnamefont {Klaric}},\ }\bibfield  {title} {\bibinfo
  {title} {{Probing Leptogenesis at Future Colliders}},\ }\href
  {https://doi.org/10.1007/JHEP09(2018)124} {\bibfield  {journal} {\bibinfo
  {journal} {JHEP}\ }\textbf {\bibinfo {volume} {09}},\ \bibinfo {pages}
  {124}},\ \Eprint {https://arxiv.org/abs/1710.03744} {arXiv:1710.03744
  [hep-ph]} \BibitemShut {NoStop}%
\bibitem [{\citenamefont {Drewes}\ \emph {et~al.}(2017)\citenamefont {Drewes},
  \citenamefont {Garbrecht}, \citenamefont {Gueter},\ and\ \citenamefont
  {Klaric}}]{Drewes:2016jae}%
  \BibitemOpen
  \bibfield  {author} {\bibinfo {author} {\bibfnamefont {M.}~\bibnamefont
  {Drewes}}, \bibinfo {author} {\bibfnamefont {B.}~\bibnamefont {Garbrecht}},
  \bibinfo {author} {\bibfnamefont {D.}~\bibnamefont {Gueter}},\ and\ \bibinfo
  {author} {\bibfnamefont {J.}~\bibnamefont {Klaric}},\ }\bibfield  {title}
  {\bibinfo {title} {{Testing the low scale seesaw and leptogenesis}},\ }\href
  {https://doi.org/10.1007/JHEP08(2017)018} {\bibfield  {journal} {\bibinfo
  {journal} {JHEP}\ }\textbf {\bibinfo {volume} {08}},\ \bibinfo {pages}
  {018}},\ \Eprint {https://arxiv.org/abs/1609.09069} {arXiv:1609.09069
  [hep-ph]} \BibitemShut {NoStop}%
\bibitem [{\citenamefont {Aad}\ \emph {et~al.}(2024)\citenamefont {Aad} \emph
  {et~al.}}]{ATLAS:2024ish}%
  \BibitemOpen
  \bibfield  {author} {\bibinfo {author} {\bibfnamefont {G.}~\bibnamefont
  {Aad}} \emph {et~al.} (\bibinfo {collaboration} {ATLAS}),\ }\bibfield
  {title} {\bibinfo {title} {{Combination of Searches for Higgs Boson Pair
  Production in pp Collisions at s=13{\,}{\,}TeV with the ATLAS Detector}},\
  }\href {https://doi.org/10.1103/PhysRevLett.133.101801} {\bibfield  {journal}
  {\bibinfo  {journal} {Phys. Rev. Lett.}\ }\textbf {\bibinfo {volume} {133}},\
  \bibinfo {pages} {101801} (\bibinfo {year} {2024})},\ \Eprint
  {https://arxiv.org/abs/2406.09971} {arXiv:2406.09971 [hep-ex]} \BibitemShut
  {NoStop}%
\bibitem [{\citenamefont {Hayrapetyan}\ \emph {et~al.}(2025)\citenamefont
  {Hayrapetyan} \emph {et~al.}}]{CMS:2024awa}%
  \BibitemOpen
  \bibfield  {author} {\bibinfo {author} {\bibfnamefont {A.}~\bibnamefont
  {Hayrapetyan}} \emph {et~al.} (\bibinfo {collaboration} {CMS}),\ }\bibfield
  {title} {\bibinfo {title} {{Constraints on the Higgs boson self-coupling from
  the combination of single and double Higgs boson production in proton-proton
  collisions at s=13TeV}},\ }\href
  {https://doi.org/10.1016/j.physletb.2024.139210} {\bibfield  {journal}
  {\bibinfo  {journal} {Phys. Lett. B}\ }\textbf {\bibinfo {volume} {861}},\
  \bibinfo {pages} {139210} (\bibinfo {year} {2025})},\ \Eprint
  {https://arxiv.org/abs/2407.13554} {arXiv:2407.13554 [hep-ex]} \BibitemShut
  {NoStop}%
\bibitem [{\citenamefont {Goertz}\ \emph {et~al.}(2013)\citenamefont {Goertz},
  \citenamefont {Papaefstathiou}, \citenamefont {Yang},\ and\ \citenamefont
  {Zurita}}]{Goertz:2013kp}%
  \BibitemOpen
  \bibfield  {author} {\bibinfo {author} {\bibfnamefont {F.}~\bibnamefont
  {Goertz}}, \bibinfo {author} {\bibfnamefont {A.}~\bibnamefont
  {Papaefstathiou}}, \bibinfo {author} {\bibfnamefont {L.~L.}\ \bibnamefont
  {Yang}},\ and\ \bibinfo {author} {\bibfnamefont {J.}~\bibnamefont {Zurita}},\
  }\bibfield  {title} {\bibinfo {title} {{Higgs Boson self-coupling
  measurements using ratios of cross sections}},\ }\href
  {https://doi.org/10.1007/JHEP06(2013)016} {\bibfield  {journal} {\bibinfo
  {journal} {JHEP}\ }\textbf {\bibinfo {volume} {06}},\ \bibinfo {pages}
  {016}},\ \Eprint {https://arxiv.org/abs/1301.3492} {arXiv:1301.3492 [hep-ph]}
  \BibitemShut {NoStop}%
\bibitem [{\citenamefont {Barger}\ \emph {et~al.}(2014)\citenamefont {Barger},
  \citenamefont {Everett}, \citenamefont {Jackson},\ and\ \citenamefont
  {Shaughnessy}}]{Barger:2013jfa}%
  \BibitemOpen
  \bibfield  {author} {\bibinfo {author} {\bibfnamefont {V.}~\bibnamefont
  {Barger}}, \bibinfo {author} {\bibfnamefont {L.~L.}\ \bibnamefont {Everett}},
  \bibinfo {author} {\bibfnamefont {C.~B.}\ \bibnamefont {Jackson}},\ and\
  \bibinfo {author} {\bibfnamefont {G.}~\bibnamefont {Shaughnessy}},\
  }\bibfield  {title} {\bibinfo {title} {{Higgs-Pair Production and Measurement
  of the Triscalar Coupling at LHC(8,14)}},\ }\href
  {https://doi.org/10.1016/j.physletb.2013.12.013} {\bibfield  {journal}
  {\bibinfo  {journal} {Phys. Lett. B}\ }\textbf {\bibinfo {volume} {728}},\
  \bibinfo {pages} {433} (\bibinfo {year} {2014})},\ \Eprint
  {https://arxiv.org/abs/1311.2931} {arXiv:1311.2931 [hep-ph]} \BibitemShut
  {NoStop}%
\bibitem [{\citenamefont {Barr}\ \emph {et~al.}(2015)\citenamefont {Barr},
  \citenamefont {Dolan}, \citenamefont {Englert}, \citenamefont {Ferreira~de
  Lima},\ and\ \citenamefont {Spannowsky}}]{Barr:2014sga}%
  \BibitemOpen
  \bibfield  {author} {\bibinfo {author} {\bibfnamefont {A.~J.}\ \bibnamefont
  {Barr}}, \bibinfo {author} {\bibfnamefont {M.~J.}\ \bibnamefont {Dolan}},
  \bibinfo {author} {\bibfnamefont {C.}~\bibnamefont {Englert}}, \bibinfo
  {author} {\bibfnamefont {D.~E.}\ \bibnamefont {Ferreira~de Lima}},\ and\
  \bibinfo {author} {\bibfnamefont {M.}~\bibnamefont {Spannowsky}},\ }\bibfield
   {title} {\bibinfo {title} {{Higgs Self-Coupling Measurements at a 100 TeV
  Hadron Collider}},\ }\href {https://doi.org/10.1007/JHEP02(2015)016}
  {\bibfield  {journal} {\bibinfo  {journal} {JHEP}\ }\textbf {\bibinfo
  {volume} {02}},\ \bibinfo {pages} {016}},\ \Eprint
  {https://arxiv.org/abs/1412.7154} {arXiv:1412.7154 [hep-ph]} \BibitemShut
  {NoStop}%
\bibitem [{\citenamefont {Caprini}\ \emph {et~al.}(2016)\citenamefont {Caprini}
  \emph {et~al.}}]{Caprini:2015zlo}%
  \BibitemOpen
  \bibfield  {author} {\bibinfo {author} {\bibfnamefont {C.}~\bibnamefont
  {Caprini}} \emph {et~al.},\ }\bibfield  {title} {\bibinfo {title} {{Science
  with the space-based interferometer eLISA. II: Gravitational waves from
  cosmological phase transitions}},\ }\href
  {https://doi.org/10.1088/1475-7516/2016/04/001} {\bibfield  {journal}
  {\bibinfo  {journal} {JCAP}\ }\textbf {\bibinfo {volume} {04}},\ \bibinfo
  {pages} {001}},\ \Eprint {https://arxiv.org/abs/1512.06239} {arXiv:1512.06239
  [astro-ph.CO]} \BibitemShut {NoStop}%
\bibitem [{\citenamefont {Amaro-Seoane}\ \emph {et~al.}(2017)\citenamefont
  {Amaro-Seoane} \emph {et~al.}}]{LISA:2017pwj}%
  \BibitemOpen
  \bibfield  {author} {\bibinfo {author} {\bibfnamefont {P.}~\bibnamefont
  {Amaro-Seoane}} \emph {et~al.} (\bibinfo {collaboration} {LISA}),\ }\bibfield
   {title} {\bibinfo {title} {{Laser Interferometer Space Antenna}},\
  }\href@noop {} {\  (\bibinfo {year} {2017})},\ \Eprint
  {https://arxiv.org/abs/1702.00786} {arXiv:1702.00786 [astro-ph.IM]}
  \BibitemShut {NoStop}%
\bibitem [{\citenamefont {Yagi}\ and\ \citenamefont
  {Seto}(2011)}]{Yagi:2011wg}%
  \BibitemOpen
  \bibfield  {author} {\bibinfo {author} {\bibfnamefont {K.}~\bibnamefont
  {Yagi}}\ and\ \bibinfo {author} {\bibfnamefont {N.}~\bibnamefont {Seto}},\
  }\bibfield  {title} {\bibinfo {title} {{Detector configuration of DECIGO/BBO
  and identification of cosmological neutron-star binaries}},\ }\href
  {https://doi.org/10.1103/PhysRevD.83.044011} {\bibfield  {journal} {\bibinfo
  {journal} {Phys. Rev. D}\ }\textbf {\bibinfo {volume} {83}},\ \bibinfo
  {pages} {044011} (\bibinfo {year} {2011})},\ \bibinfo {note} {[Erratum:
  Phys.Rev.D 95, 109901 (2017)]},\ \Eprint {https://arxiv.org/abs/1101.3940}
  {arXiv:1101.3940 [astro-ph.CO]} \BibitemShut {NoStop}%
\bibitem [{\citenamefont {Crowder}\ and\ \citenamefont
  {Cornish}(2005)}]{Crowder:2005nr}%
  \BibitemOpen
  \bibfield  {author} {\bibinfo {author} {\bibfnamefont {J.}~\bibnamefont
  {Crowder}}\ and\ \bibinfo {author} {\bibfnamefont {N.~J.}\ \bibnamefont
  {Cornish}},\ }\bibfield  {title} {\bibinfo {title} {{Beyond LISA: Exploring
  future gravitational wave missions}},\ }\href
  {https://doi.org/10.1103/PhysRevD.72.083005} {\bibfield  {journal} {\bibinfo
  {journal} {Phys. Rev. D}\ }\textbf {\bibinfo {volume} {72}},\ \bibinfo
  {pages} {083005} (\bibinfo {year} {2005})},\ \Eprint
  {https://arxiv.org/abs/gr-qc/0506015} {arXiv:gr-qc/0506015} \BibitemShut
  {NoStop}%
\bibitem [{\citenamefont {Corbin}\ and\ \citenamefont
  {Cornish}(2006)}]{Corbin:2005ny}%
  \BibitemOpen
  \bibfield  {author} {\bibinfo {author} {\bibfnamefont {V.}~\bibnamefont
  {Corbin}}\ and\ \bibinfo {author} {\bibfnamefont {N.~J.}\ \bibnamefont
  {Cornish}},\ }\bibfield  {title} {\bibinfo {title} {{Detecting the cosmic
  gravitational wave background with the big bang observer}},\ }\href
  {https://doi.org/10.1088/0264-9381/23/7/014} {\bibfield  {journal} {\bibinfo
  {journal} {Class. Quant. Grav.}\ }\textbf {\bibinfo {volume} {23}},\ \bibinfo
  {pages} {2435} (\bibinfo {year} {2006})},\ \Eprint
  {https://arxiv.org/abs/gr-qc/0512039} {arXiv:gr-qc/0512039} \BibitemShut
  {NoStop}%
\bibitem [{\citenamefont {Harry}\ \emph {et~al.}(2006)\citenamefont {Harry},
  \citenamefont {Fritschel}, \citenamefont {Shaddock}, \citenamefont
  {Folkner},\ and\ \citenamefont {Phinney}}]{Harry:2006fi}%
  \BibitemOpen
  \bibfield  {author} {\bibinfo {author} {\bibfnamefont {G.~M.}\ \bibnamefont
  {Harry}}, \bibinfo {author} {\bibfnamefont {P.}~\bibnamefont {Fritschel}},
  \bibinfo {author} {\bibfnamefont {D.~A.}\ \bibnamefont {Shaddock}}, \bibinfo
  {author} {\bibfnamefont {W.}~\bibnamefont {Folkner}},\ and\ \bibinfo {author}
  {\bibfnamefont {E.~S.}\ \bibnamefont {Phinney}},\ }\bibfield  {title}
  {\bibinfo {title} {{Laser interferometry for the big bang observer}},\ }\href
  {https://doi.org/10.1088/0264-9381/23/15/008} {\bibfield  {journal} {\bibinfo
   {journal} {Class. Quant. Grav.}\ }\textbf {\bibinfo {volume} {23}},\
  \bibinfo {pages} {4887} (\bibinfo {year} {2006})},\ \bibinfo {note}
  {[Erratum: Class.Quant.Grav. 23, 7361 (2006)]}\BibitemShut {NoStop}%
\bibitem [{\citenamefont {Kawamura}\ \emph {et~al.}(2006)\citenamefont
  {Kawamura} \emph {et~al.}}]{Kawamura:2006up}%
  \BibitemOpen
  \bibfield  {author} {\bibinfo {author} {\bibfnamefont {S.}~\bibnamefont
  {Kawamura}} \emph {et~al.},\ }\bibfield  {title} {\bibinfo {title} {{The
  Japanese space gravitational wave antenna DECIGO}},\ }\href
  {https://doi.org/10.1088/0264-9381/23/8/S17} {\bibfield  {journal} {\bibinfo
  {journal} {Class. Quant. Grav.}\ }\textbf {\bibinfo {volume} {23}},\ \bibinfo
  {pages} {S125} (\bibinfo {year} {2006})}\BibitemShut {NoStop}%
\bibitem [{\citenamefont {Sesana}\ \emph {et~al.}(2021)\citenamefont {Sesana}
  \emph {et~al.}}]{Sesana:2019vho}%
  \BibitemOpen
  \bibfield  {author} {\bibinfo {author} {\bibfnamefont {A.}~\bibnamefont
  {Sesana}} \emph {et~al.},\ }\bibfield  {title} {\bibinfo {title} {{Unveiling
  the gravitational universe at $\mu$-Hz frequencies}},\ }\href
  {https://doi.org/10.1007/s10686-021-09709-9} {\bibfield  {journal} {\bibinfo
  {journal} {Exper. Astron.}\ }\textbf {\bibinfo {volume} {51}},\ \bibinfo
  {pages} {1333} (\bibinfo {year} {2021})},\ \Eprint
  {https://arxiv.org/abs/1908.11391} {arXiv:1908.11391 [astro-ph.IM]}
  \BibitemShut {NoStop}%
\bibitem [{\citenamefont {Abbott}\ \emph {et~al.}(2017)\citenamefont {Abbott}
  \emph {et~al.}}]{LIGOScientific:2016wof}%
  \BibitemOpen
  \bibfield  {author} {\bibinfo {author} {\bibfnamefont {B.~P.}\ \bibnamefont
  {Abbott}} \emph {et~al.} (\bibinfo {collaboration} {LIGO Scientific}),\
  }\bibfield  {title} {\bibinfo {title} {{Exploring the Sensitivity of Next
  Generation Gravitational Wave Detectors}},\ }\href
  {https://doi.org/10.1088/1361-6382/aa51f4} {\bibfield  {journal} {\bibinfo
  {journal} {Class. Quant. Grav.}\ }\textbf {\bibinfo {volume} {34}},\ \bibinfo
  {pages} {044001} (\bibinfo {year} {2017})},\ \Eprint
  {https://arxiv.org/abs/1607.08697} {arXiv:1607.08697 [astro-ph.IM]}
  \BibitemShut {NoStop}%
\bibitem [{\citenamefont {Reitze}\ \emph {et~al.}(2019)\citenamefont {Reitze}
  \emph {et~al.}}]{Reitze:2019iox}%
  \BibitemOpen
  \bibfield  {author} {\bibinfo {author} {\bibfnamefont {D.}~\bibnamefont
  {Reitze}} \emph {et~al.},\ }\bibfield  {title} {\bibinfo {title} {{Cosmic
  Explorer: The U.S. Contribution to Gravitational-Wave Astronomy beyond
  LIGO}},\ }\href@noop {} {\bibfield  {journal} {\bibinfo  {journal} {Bull. Am.
  Astron. Soc.}\ }\textbf {\bibinfo {volume} {51}},\ \bibinfo {pages} {035}
  (\bibinfo {year} {2019})},\ \Eprint {https://arxiv.org/abs/1907.04833}
  {arXiv:1907.04833 [astro-ph.IM]} \BibitemShut {NoStop}%
\bibitem [{\citenamefont {Punturo}\ \emph {et~al.}(2010)\citenamefont {Punturo}
  \emph {et~al.}}]{Punturo:2010zz}%
  \BibitemOpen
  \bibfield  {author} {\bibinfo {author} {\bibfnamefont {M.}~\bibnamefont
  {Punturo}} \emph {et~al.},\ }\bibfield  {title} {\bibinfo {title} {{The
  Einstein Telescope: A third-generation gravitational wave observatory}},\
  }\href {https://doi.org/10.1088/0264-9381/27/19/194002} {\bibfield  {journal}
  {\bibinfo  {journal} {Class. Quant. Grav.}\ }\textbf {\bibinfo {volume}
  {27}},\ \bibinfo {pages} {194002} (\bibinfo {year} {2010})}\BibitemShut
  {NoStop}%
\bibitem [{\citenamefont {Hild}\ \emph {et~al.}(2011)\citenamefont {Hild} \emph
  {et~al.}}]{Hild:2010id}%
  \BibitemOpen
  \bibfield  {author} {\bibinfo {author} {\bibfnamefont {S.}~\bibnamefont
  {Hild}} \emph {et~al.},\ }\bibfield  {title} {\bibinfo {title} {{Sensitivity
  Studies for Third-Generation Gravitational Wave Observatories}},\ }\href
  {https://doi.org/10.1088/0264-9381/28/9/094013} {\bibfield  {journal}
  {\bibinfo  {journal} {Class. Quant. Grav.}\ }\textbf {\bibinfo {volume}
  {28}},\ \bibinfo {pages} {094013} (\bibinfo {year} {2011})},\ \Eprint
  {https://arxiv.org/abs/1012.0908} {arXiv:1012.0908 [gr-qc]} \BibitemShut
  {NoStop}%
\bibitem [{\citenamefont {Sathyaprakash}\ \emph {et~al.}(2012)\citenamefont
  {Sathyaprakash} \emph {et~al.}}]{Sathyaprakash:2012jk}%
  \BibitemOpen
  \bibfield  {author} {\bibinfo {author} {\bibfnamefont {B.}~\bibnamefont
  {Sathyaprakash}} \emph {et~al.},\ }\bibfield  {title} {\bibinfo {title}
  {{Scientific Objectives of Einstein Telescope}},\ }\href
  {https://doi.org/10.1088/0264-9381/29/12/124013} {\bibfield  {journal}
  {\bibinfo  {journal} {Class. Quant. Grav.}\ }\textbf {\bibinfo {volume}
  {29}},\ \bibinfo {pages} {124013} (\bibinfo {year} {2012})},\ \bibinfo {note}
  {[Erratum: Class.Quant.Grav. 30, 079501 (2013)]},\ \Eprint
  {https://arxiv.org/abs/1206.0331} {arXiv:1206.0331 [gr-qc]} \BibitemShut
  {NoStop}%
\bibitem [{\citenamefont {Maggiore}\ \emph {et~al.}(2020)\citenamefont
  {Maggiore} \emph {et~al.}}]{ET:2019dnz}%
  \BibitemOpen
  \bibfield  {author} {\bibinfo {author} {\bibfnamefont {M.}~\bibnamefont
  {Maggiore}} \emph {et~al.} (\bibinfo {collaboration} {ET}),\ }\bibfield
  {title} {\bibinfo {title} {{Science Case for the Einstein Telescope}},\
  }\href {https://doi.org/10.1088/1475-7516/2020/03/050} {\bibfield  {journal}
  {\bibinfo  {journal} {JCAP}\ }\textbf {\bibinfo {volume} {03}},\ \bibinfo
  {pages} {050}},\ \Eprint {https://arxiv.org/abs/1912.02622} {arXiv:1912.02622
  [astro-ph.CO]} \BibitemShut {NoStop}%
\bibitem [{\citenamefont {Garcia-Bellido}\ \emph {et~al.}(2021)\citenamefont
  {Garcia-Bellido}, \citenamefont {Murayama},\ and\ \citenamefont
  {White}}]{Garcia-Bellido:2021zgu}%
  \BibitemOpen
  \bibfield  {author} {\bibinfo {author} {\bibfnamefont {J.}~\bibnamefont
  {Garcia-Bellido}}, \bibinfo {author} {\bibfnamefont {H.}~\bibnamefont
  {Murayama}},\ and\ \bibinfo {author} {\bibfnamefont {G.}~\bibnamefont
  {White}},\ }\bibfield  {title} {\bibinfo {title} {{Exploring the early
  Universe with Gaia and Theia}},\ }\href
  {https://doi.org/10.1088/1475-7516/2021/12/023} {\bibfield  {journal}
  {\bibinfo  {journal} {JCAP}\ }\textbf {\bibinfo {volume} {12}}\bibfield
  {number} {\bibinfo  {number} { (12)},\ \bibinfo {pages} {023}},\ }\Eprint
  {https://arxiv.org/abs/2104.04778} {arXiv:2104.04778 [hep-ph]} \BibitemShut
  {NoStop}%
\bibitem [{\citenamefont {Coleman}\ and\ \citenamefont
  {Weinberg}(1973)}]{Coleman:1973jx}%
  \BibitemOpen
  \bibfield  {author} {\bibinfo {author} {\bibfnamefont {S.~R.}\ \bibnamefont
  {Coleman}}\ and\ \bibinfo {author} {\bibfnamefont {E.~J.}\ \bibnamefont
  {Weinberg}},\ }\bibfield  {title} {\bibinfo {title} {{Radiative Corrections
  as the Origin of Spontaneous Symmetry Breaking}},\ }\href
  {https://doi.org/10.1103/PhysRevD.7.1888} {\bibfield  {journal} {\bibinfo
  {journal} {Phys. Rev. D}\ }\textbf {\bibinfo {volume} {7}},\ \bibinfo {pages}
  {1888} (\bibinfo {year} {1973})}\BibitemShut {NoStop}%
\bibitem [{\citenamefont {Quiros}(1999)}]{Quiros:1999jp}%
  \BibitemOpen
  \bibfield  {author} {\bibinfo {author} {\bibfnamefont {M.}~\bibnamefont
  {Quiros}},\ }\bibfield  {title} {\bibinfo {title} {{Finite temperature field
  theory and phase transitions}},\ }in\ \href@noop {} {\emph {\bibinfo
  {booktitle} {{ICTP Summer School in High-Energy Physics and Cosmology}}}}\
  (\bibinfo {year} {1999})\ pp.\ \bibinfo {pages} {187--259},\ \Eprint
  {https://arxiv.org/abs/hep-ph/9901312} {arXiv:hep-ph/9901312} \BibitemShut
  {NoStop}%
\bibitem [{\citenamefont {Arnold}\ and\ \citenamefont
  {Espinosa}(1993)}]{Arnold:1992rz}%
  \BibitemOpen
  \bibfield  {author} {\bibinfo {author} {\bibfnamefont {P.~B.}\ \bibnamefont
  {Arnold}}\ and\ \bibinfo {author} {\bibfnamefont {O.}~\bibnamefont
  {Espinosa}},\ }\bibfield  {title} {\bibinfo {title} {{The Effective potential
  and first order phase transitions: Beyond leading-order}},\ }\href
  {https://doi.org/10.1103/PhysRevD.47.3546} {\bibfield  {journal} {\bibinfo
  {journal} {Phys. Rev. D}\ }\textbf {\bibinfo {volume} {47}},\ \bibinfo
  {pages} {3546} (\bibinfo {year} {1993})},\ \bibinfo {note} {[Erratum:
  Phys.Rev.D 50, 6662 (1994)]},\ \Eprint {https://arxiv.org/abs/hep-ph/9212235}
  {arXiv:hep-ph/9212235} \BibitemShut {NoStop}%
\bibitem [{\citenamefont {Carrington}(1992)}]{Carrington:1991hz}%
  \BibitemOpen
  \bibfield  {author} {\bibinfo {author} {\bibfnamefont {M.~E.}\ \bibnamefont
  {Carrington}},\ }\bibfield  {title} {\bibinfo {title} {{The Effective
  potential at finite temperature in the Standard Model}},\ }\href
  {https://doi.org/10.1103/PhysRevD.45.2933} {\bibfield  {journal} {\bibinfo
  {journal} {Phys. Rev. D}\ }\textbf {\bibinfo {volume} {45}},\ \bibinfo
  {pages} {2933} (\bibinfo {year} {1992})}\BibitemShut {NoStop}%
\bibitem [{\citenamefont {Parwani}(1992)}]{Parwani:1991gq}%
  \BibitemOpen
  \bibfield  {author} {\bibinfo {author} {\bibfnamefont {R.~R.}\ \bibnamefont
  {Parwani}},\ }\bibfield  {title} {\bibinfo {title} {{Resummation in a hot
  scalar field theory}},\ }\href {https://doi.org/10.1103/PhysRevD.45.4695}
  {\bibfield  {journal} {\bibinfo  {journal} {Phys. Rev. D}\ }\textbf {\bibinfo
  {volume} {45}},\ \bibinfo {pages} {4695} (\bibinfo {year} {1992})},\ \bibinfo
  {note} {[Erratum: Phys.Rev.D 48, 5965 (1993)]},\ \Eprint
  {https://arxiv.org/abs/hep-ph/9204216} {arXiv:hep-ph/9204216} \BibitemShut
  {NoStop}%
\bibitem [{\citenamefont {Athron}\ \emph {et~al.}(2024)\citenamefont {Athron},
  \citenamefont {Bal\'azs}, \citenamefont {Fowlie}, \citenamefont {Morris},\
  and\ \citenamefont {Wu}}]{Athron:2023xlk}%
  \BibitemOpen
  \bibfield  {author} {\bibinfo {author} {\bibfnamefont {P.}~\bibnamefont
  {Athron}}, \bibinfo {author} {\bibfnamefont {C.}~\bibnamefont {Bal\'azs}},
  \bibinfo {author} {\bibfnamefont {A.}~\bibnamefont {Fowlie}}, \bibinfo
  {author} {\bibfnamefont {L.}~\bibnamefont {Morris}},\ and\ \bibinfo {author}
  {\bibfnamefont {L.}~\bibnamefont {Wu}},\ }\bibfield  {title} {\bibinfo
  {title} {{Cosmological phase transitions: From perturbative particle physics
  to gravitational waves}},\ }\href
  {https://doi.org/10.1016/j.ppnp.2023.104094} {\bibfield  {journal} {\bibinfo
  {journal} {Prog. Part. Nucl. Phys.}\ }\textbf {\bibinfo {volume} {135}},\
  \bibinfo {pages} {104094} (\bibinfo {year} {2024})},\ \Eprint
  {https://arxiv.org/abs/2305.02357} {arXiv:2305.02357 [hep-ph]} \BibitemShut
  {NoStop}%
\bibitem [{\citenamefont {Guo}\ \emph {et~al.}(2021)\citenamefont {Guo},
  \citenamefont {Sinha}, \citenamefont {Vagie},\ and\ \citenamefont
  {White}}]{Guo:2020grp}%
  \BibitemOpen
  \bibfield  {author} {\bibinfo {author} {\bibfnamefont {H.-K.}\ \bibnamefont
  {Guo}}, \bibinfo {author} {\bibfnamefont {K.}~\bibnamefont {Sinha}}, \bibinfo
  {author} {\bibfnamefont {D.}~\bibnamefont {Vagie}},\ and\ \bibinfo {author}
  {\bibfnamefont {G.}~\bibnamefont {White}},\ }\bibfield  {title} {\bibinfo
  {title} {{Phase Transitions in an Expanding Universe: Stochastic
  Gravitational Waves in Standard and Non-Standard Histories}},\ }\href
  {https://doi.org/10.1088/1475-7516/2021/01/001} {\bibfield  {journal}
  {\bibinfo  {journal} {JCAP}\ }\textbf {\bibinfo {volume} {01}},\ \bibinfo
  {pages} {001}},\ \Eprint {https://arxiv.org/abs/2007.08537} {arXiv:2007.08537
  [hep-ph]} \BibitemShut {NoStop}%
\bibitem [{\citenamefont {Caprini}\ \emph {et~al.}(2020)\citenamefont {Caprini}
  \emph {et~al.}}]{Caprini:2019egz}%
  \BibitemOpen
  \bibfield  {author} {\bibinfo {author} {\bibfnamefont {C.}~\bibnamefont
  {Caprini}} \emph {et~al.},\ }\bibfield  {title} {\bibinfo {title} {{Detecting
  gravitational waves from cosmological phase transitions with LISA: an
  update}},\ }\href {https://doi.org/10.1088/1475-7516/2020/03/024} {\bibfield
  {journal} {\bibinfo  {journal} {JCAP}\ }\textbf {\bibinfo {volume} {03}},\
  \bibinfo {pages} {024}},\ \Eprint {https://arxiv.org/abs/1910.13125}
  {arXiv:1910.13125 [astro-ph.CO]} \BibitemShut {NoStop}%
\bibitem [{\citenamefont {Hindmarsh}\ \emph {et~al.}(2021)\citenamefont
  {Hindmarsh}, \citenamefont {L\"uben}, \citenamefont {Lumma},\ and\
  \citenamefont {Pauly}}]{Hindmarsh:2020hop}%
  \BibitemOpen
  \bibfield  {author} {\bibinfo {author} {\bibfnamefont {M.~B.}\ \bibnamefont
  {Hindmarsh}}, \bibinfo {author} {\bibfnamefont {M.}~\bibnamefont {L\"uben}},
  \bibinfo {author} {\bibfnamefont {J.}~\bibnamefont {Lumma}},\ and\ \bibinfo
  {author} {\bibfnamefont {M.}~\bibnamefont {Pauly}},\ }\bibfield  {title}
  {\bibinfo {title} {{Phase transitions in the early universe}},\ }\href
  {https://doi.org/10.21468/SciPostPhysLectNotes.24} {\bibfield  {journal}
  {\bibinfo  {journal} {SciPost Phys. Lect. Notes}\ }\textbf {\bibinfo {volume}
  {24}},\ \bibinfo {pages} {1} (\bibinfo {year} {2021})},\ \Eprint
  {https://arxiv.org/abs/2008.09136} {arXiv:2008.09136 [astro-ph.CO]}
  \BibitemShut {NoStop}%
\bibitem [{\citenamefont {Grojean}\ and\ \citenamefont
  {Servant}(2007)}]{Grojean:2006bp}%
  \BibitemOpen
  \bibfield  {author} {\bibinfo {author} {\bibfnamefont {C.}~\bibnamefont
  {Grojean}}\ and\ \bibinfo {author} {\bibfnamefont {G.}~\bibnamefont
  {Servant}},\ }\bibfield  {title} {\bibinfo {title} {{Gravitational Waves from
  Phase Transitions at the Electroweak Scale and Beyond}},\ }\href
  {https://doi.org/10.1103/PhysRevD.75.043507} {\bibfield  {journal} {\bibinfo
  {journal} {Phys. Rev. D}\ }\textbf {\bibinfo {volume} {75}},\ \bibinfo
  {pages} {043507} (\bibinfo {year} {2007})},\ \Eprint
  {https://arxiv.org/abs/hep-ph/0607107} {arXiv:hep-ph/0607107} \BibitemShut
  {NoStop}%
\bibitem [{\citenamefont {Vaskonen}(2017)}]{Vaskonen:2016yiu}%
  \BibitemOpen
  \bibfield  {author} {\bibinfo {author} {\bibfnamefont {V.}~\bibnamefont
  {Vaskonen}},\ }\bibfield  {title} {\bibinfo {title} {{Electroweak
  baryogenesis and gravitational waves from a real scalar singlet}},\ }\href
  {https://doi.org/10.1103/PhysRevD.95.123515} {\bibfield  {journal} {\bibinfo
  {journal} {Phys. Rev. D}\ }\textbf {\bibinfo {volume} {95}},\ \bibinfo
  {pages} {123515} (\bibinfo {year} {2017})},\ \Eprint
  {https://arxiv.org/abs/1611.02073} {arXiv:1611.02073 [hep-ph]} \BibitemShut
  {NoStop}%
\bibitem [{\citenamefont {Espinosa}\ \emph {et~al.}(2010)\citenamefont
  {Espinosa}, \citenamefont {Konstandin}, \citenamefont {No},\ and\
  \citenamefont {Servant}}]{Espinosa:2010hh}%
  \BibitemOpen
  \bibfield  {author} {\bibinfo {author} {\bibfnamefont {J.~R.}\ \bibnamefont
  {Espinosa}}, \bibinfo {author} {\bibfnamefont {T.}~\bibnamefont
  {Konstandin}}, \bibinfo {author} {\bibfnamefont {J.~M.}\ \bibnamefont {No}},\
  and\ \bibinfo {author} {\bibfnamefont {G.}~\bibnamefont {Servant}},\
  }\bibfield  {title} {\bibinfo {title} {{Energy Budget of Cosmological
  First-order Phase Transitions}},\ }\href
  {https://doi.org/10.1088/1475-7516/2010/06/028} {\bibfield  {journal}
  {\bibinfo  {journal} {JCAP}\ }\textbf {\bibinfo {volume} {06}},\ \bibinfo
  {pages} {028}},\ \Eprint {https://arxiv.org/abs/1004.4187} {arXiv:1004.4187
  [hep-ph]} \BibitemShut {NoStop}%
\bibitem [{\citenamefont {Hindmarsh}\ \emph {et~al.}(2015)\citenamefont
  {Hindmarsh}, \citenamefont {Huber}, \citenamefont {Rummukainen},\ and\
  \citenamefont {Weir}}]{Hindmarsh:2015qta}%
  \BibitemOpen
  \bibfield  {author} {\bibinfo {author} {\bibfnamefont {M.}~\bibnamefont
  {Hindmarsh}}, \bibinfo {author} {\bibfnamefont {S.~J.}\ \bibnamefont
  {Huber}}, \bibinfo {author} {\bibfnamefont {K.}~\bibnamefont {Rummukainen}},\
  and\ \bibinfo {author} {\bibfnamefont {D.~J.}\ \bibnamefont {Weir}},\
  }\bibfield  {title} {\bibinfo {title} {{Numerical simulations of acoustically
  generated gravitational waves at a first order phase transition}},\ }\href
  {https://doi.org/10.1103/PhysRevD.92.123009} {\bibfield  {journal} {\bibinfo
  {journal} {Phys. Rev. D}\ }\textbf {\bibinfo {volume} {92}},\ \bibinfo
  {pages} {123009} (\bibinfo {year} {2015})},\ \Eprint
  {https://arxiv.org/abs/1504.03291} {arXiv:1504.03291 [astro-ph.CO]}
  \BibitemShut {NoStop}%
\end{thebibliography}%

\end{document}